\documentclass[twocolumn,floatfix]{aastex631}

\newcommand{\codesnip}[1]{\texttt{#1}}

\usepackage[]{hyperref}
\hypersetup{
  colorlinks   = true, 
  urlcolor     = blue,
  linkcolor    = blue, 
  citecolor   = blue 
}
\usepackage{amsmath}	

\DeclareMathOperator*{\argmin}{argmin}

\graphicspath{{./}{figures/}}

\begin{document}
\title[Non-parametric linear models]{exoALMA. VIII. Probabilistic Moment Maps and Data Products using Non-parametric Linear Models}



\author[0000-0001-7641-5235]{Thomas Hilder}
\affiliation{School of Physics and Astronomy, Monash University, VIC 3800, Australia}

\author[0000-0003-0174-0564]{Andrew R. Casey}
\affiliation{School of Physics and Astronomy, Monash University, VIC 3800, Australia}
\affiliation{Center for Computational Astrophysics, Flatiron Institute, 162 5th Avenue, New York}
\affiliation{Centre of Excellence for Astrophysics in Three Dimensions (ASTRO-3D), Melbourne, Victoria, Australia}

\author[0000-0002-4716-4235]{Daniel J. Price}     
\affiliation{School of Physics and Astronomy, Monash University, VIC 3800, Australia}

\author[0000-0001-5907-5179]{Christophe Pinte}
\affiliation{Univ. Grenoble Alpes, CNRS, IPAG, 38000 Grenoble, France}
\affiliation{School of Physics and Astronomy, Monash University, VIC 3800, Australia}

\author[0000-0001-8446-3026]{Andr\'es F. Izquierdo}
\altaffiliation{NASA Hubble Fellowship Program Sagan Fellow}
\affiliation{Department of Astronomy, University of Florida, Gainesville, FL 32611, USA}
\affiliation{Leiden Observatory, Leiden University, P.O. Box 9513, NL-2300 RA Leiden, The Netherlands}
\affiliation{European Southern Observatory, Karl-Schwarzschild-Str. 2, D-85748 Garching bei M\"nchen, Germany}

\author[0009-0003-7403-9207]{Caitlyn Hardiman}
\affiliation{School of Physics and Astronomy, Monash University, VIC 3800, Australia}


\author[0000-0001-7258-770X]{Jaehan Bae}
\affiliation{Department of Astronomy, University of Florida, Gainesville, FL 32611, USA}

\author[0000-0001-6378-7873]{Marcelo Barraza-Alfaro}
\affiliation{Department of Earth, Atmospheric, and Planetary Sciences, Massachusetts Institute of Technology, Cambridge, MA 02139, USA}

\author[0000-0002-7695-7605]{Myriam Benisty}
\affiliation{Universit\'{e} C\^{o}te d'Azur, Observatoire de la C\^{o}te d'Azur, CNRS, Laboratoire Lagrange, France}
\affiliation{Max-Planck Institute for Astronomy (MPIA), Königstuhl 17, 69117 Heidelberg, Germany}

\author[0000-0002-2700-9676]{Gianni Cataldi}
\affiliation{National Astronomical Observatory of Japan, 2-21-1 Osawa, Mitaka, Tokyo 181-8588, Japan}

\author[0000-0003-2045-2154]{Pietro Curone}
\affiliation{Dipartimento di Fisica, Universit\`a degli Studi di Milano, Via Celoria 16, 20133 Milano, Italy}
\affiliation{Departamento de Astronomía, Universidad de Chile, Camino El Observatorio 1515, Las Condes, Santiago, Chile}

\author[0000-0002-1483-8811]{Ian Czekala}
\affiliation{School of Physics \& Astronomy, University of St. Andrews, North Haugh, St. Andrews KY16 9SS, UK}

\author[0000-0003-4689-2684]{Stefano Facchini}
\affiliation{Dipartimento di Fisica, Universit\`a degli Studi di Milano, Via Celoria 16, 20133 Milano, Italy}

\author[0000-0003-4679-4072]{Daniele Fasano}
\affiliation{Universit\'{e} C\^{o}te d'Azur, Observatoire de la C\^{o}te d'Azur, CNRS, Laboratoire Lagrange, France}

\author[0000-0002-9298-3029]{Mario Flock}
\affiliation{Max-Planck Institute for Astronomy (MPIA), Königstuhl 17, 69117 Heidelberg, Germany}

\author[0000-0003-1117-9213]{Misato Fukagawa}
\affiliation{National Astronomical Observatory of Japan, 2-21-1 Osawa, Mitaka, Tokyo 181-8588, Japan}

\author[0000-0002-5503-5476]{Maria Galloway-Sprietsma}
\affiliation{Department of Astronomy, University of Florida, Gainesville, FL 32611, USA}

\author[0000-0002-5910-4598]{Himanshi Garg}
\affiliation{School of Physics and Astronomy, Monash University, VIC 3800, Australia}

\author[0000-0002-8138-0425]{Cassandra Hall}
\affiliation{Department of Physics and Astronomy, The University of Georgia, Athens, GA 30602, USA}
\affiliation{Center for Simulational Physics, The University of Georgia, Athens, GA 30602, USA}
\affiliation{Institute for Artificial Intelligence, The University of Georgia, Athens, GA, 30602, USA}

\author[0000-0003-1502-4315]{Iain Hammond}
\affiliation{School of Physics and Astronomy, Monash University, VIC 3800, Australia}

\author[0000-0001-6947-6072]{Jane Huang}
\affiliation{Department of Astronomy, Columbia University, 538 W. 120th Street, Pupin Hall, New York, NY 10027, USA}

\author[0000-0003-1008-1142]{John D. Ilee}
\affiliation{School of Physics and Astronomy, University of Leeds, Leeds, UK, LS2 9JT}


\author[0000-0001-7235-2417]{Kazuhiro Kanagawa}
\affiliation{College of Science, Ibaraki University, 2-1-1 Bunkyo, Mito, Ibaraki 310-8512, Japan}

\author[0000-0002-8896-9435]{Geoffroy Lesur}
\affiliation{Univ. Grenoble Alpes, CNRS, IPAG, 38000 Grenoble, France}

\author[0000-0003-4663-0318]{Cristiano Longarini}
\affiliation{Institute of Astronomy, University of Cambridge, Madingley Rd, CB30HA, Cambridge, UK}
\affiliation{Dipartimento di Fisica, Universit\`a degli Studi di Milano, Via Celoria 16, 20133 Milano, Italy}

\author[0000-0002-8932-1219]{Ryan Loomis}
\affiliation{National Radio Astronomy Observatory, Charlottesville, VA 22903, USA}


\author[0000-0003-4039-8933]{Ryuta Orihara}
\affiliation{College of Science, Ibaraki University, 2-1-1 Bunkyo, Mito, Ibaraki 310-8512, Japan}



\author[0000-0003-4853-5736]{Giovanni Rosotti}
\affiliation{Dipartimento di Fisica, Universit\`a degli Studi di Milano, Via Celoria 16, 20133 Milano, Italy}

\author[0000-0002-0491-143X]{Jochen Stadler}
\affiliation{Universit\'{e} C\^{o}te d'Azur, Observatoire de la C\^{o}te d'Azur, CNRS, Laboratoire Lagrange, France}
\affiliation{Univ. Grenoble Alpes, CNRS, IPAG, 38000 Grenoble, France}

\author[0000-0003-1534-5186]{Richard Teague}
\affiliation{Department of Earth, Atmospheric, and Planetary Sciences, Massachusetts Institute of Technology, Cambridge, MA 02139, USA}

\author[0000-0003-1412-893X]{Hsi-Wei Yen}
\affiliation{Academia Sinica Institute of Astronomy \& Astrophysics, 11F of Astronomy-Mathematics Building, AS/NTU, No.1, Sec. 4, Roosevelt Rd, Taipei 10617, Taiwan}

\author[0000-0002-3468-9577]{Gaylor Wafflard}
\affiliation{Univ. Grenoble Alpes, CNRS, IPAG, 38000 Grenoble, France}

\author[0000-0002-7501-9801]{Andrew J. Winter}
\affiliation{Universit\'{e} C\^{o}te d'Azur, Observatoire de la C\^{o}te d'Azur, CNRS, Laboratoire Lagrange, France}
\affiliation{Max-Planck Institute for Astronomy (MPIA), Königstuhl 17, 69117 Heidelberg, Germany}

\author[0000-0002-7212-2416]{Lisa W\"olfer}
\affiliation{Department of Earth, Atmospheric, and Planetary Sciences, Massachusetts Institute of Technology, Cambridge, MA 02139, USA}

\author[0000-0001-8002-8473]{Tomohiro C. Yoshida}
\affiliation{National Astronomical Observatory of Japan, 2-21-1 Osawa, Mitaka, Tokyo 181-8588, Japan}
\affiliation{Department of Astronomical Science, The Graduate University for Advanced Studies, SOKENDAI, 2-21-1 Osawa, Mitaka, Tokyo 181-8588, Japan}

\author[0000-0001-9319-1296]{Brianna Zawadzki}
\affiliation{Department of Astronomy, Van Vleck Observatory, Wesleyan University, 96 Foss Hill Drive, Middletown, CT 06459, USA}
\affiliation{Department of Astronomy \& Astrophysics, 525 Davey Laboratory, The Pennsylvania State University, University Park, PA 16802, USA}

\correspondingauthor{Thomas Hilder}
\email{thomas.hilder@monash.edu}


\begin{abstract}

Extracting robust inferences on physical quantities from disk kinematics measured from Doppler-shifted molecular line emission is challenging due to the data's size and complexity.
In this paper we develop a flexible linear model of the intensity distribution in each frequency channel, accounting for spatial correlations from the point spread function.
The analytic form of the model's posterior enables probabilistic data products through sampling. 
Our method debiases peak intensity, peak velocity, and line width maps, particularly in disk substructures that are only partially resolved. 
These are needed in order to measure disk mass, turbulence, pressure gradients, and to detect embedded planets. 
We analyse HD~135344B, MWC~758, and CQ~Tau, finding velocity substructures 50--200 ${\rm m s^{-1}}$ greater than with conventional methods. 
Additionally, we combine our approach with \textsc{discminer} in a case study of J1842.
We find that uncertainties in stellar mass and inclination increase by an order of magnitude due to the more realistic noise model.  
More broadly, our method can be applied to any problem requiring a probabilistic model of an intensity distribution conditioned on a point spread function.

\end{abstract}

\section{Introduction} \label{sec:intro}

Observations of circumstellar discs in Doppler-shifted molecular line emission at high spatial and spectral resolution, as obtained in the exoALMA project \citep{Teague_exoALMA}, offer a wealth of information about the 3D velocity field \citep{Izquierdo_exoALMA}, emitting-layer height, temperature structure \citep{Galloway_exoALMA}, rotation curve, star and disk masses \citep{Longarini_exoALMA}, radial kinematic substructures \citep{Stadler_exoALMA}, turbulence \citep{Barraza_exoALMA,Hardiman_exoALMA}, and spiral-arms due to planets or stellar-mass companions \citep{Pinte_exoALMA}.

Performing these measurements in a statistically robust way is difficult.
First, the data are complicated.
At the high spatial resolutions achieved in exoALMA, we resolve many local non-axisymmetric features in both intensity and velocity \citep{Izquierdo_exoALMA}, which cannot be explained with global axisymmetric, Keplerian disk models.
While Bayesian approaches modelling either channel maps \citep{izquierdo2021,izquierdo2022,izquierdo2023} or visibilities \citep{czekala2015,flaherty2015,flaherty2020,pegues2021,kurtovic2024,Hardiman_exoALMA} with parametric disk models are able to infer all parameters simultaneously, their lack of flexibility limits the interpretability of the posterior \citep[e.g.][]{bernardo2009bayesian} when applied to discs containing substructures as seen in all of the exoALMA targets \citep{diskdynamicscollaboration2020,Teague_exoALMA,Pinte_exoALMA}. 

Second, the imaging process creates correlations between image pixels and the finite sampling of the $uv$-plane \citep[e.g.][]{andrews2018} means that the data contains no information about particular frequencies.
The resultant image point spread function results in the underestimation of line intensity gradients known as beam smearing \citep{cotton1989}. Beam smearing is known to bias line widths in the inner disk \citep{teague2016} and velocities in the presence of intensity gradients, including the local gradients caused by substructures \citep{keppler2019, boehler2021}. It also causes bias in rotation curves even at radii greater than 4 times the full width half maximum of the beam \citep{andrews2024,Stadler_exoALMA}.
Recently, \citet{andrews2024} found empirical corrections for this effect for rotation curve analysis, but such corrections are limited to axisymmetric, Keplerian discs and local intensity gradients are not accounted for.

Third, analyses looking to measure deviations from Keplerian rotation such as from pressure gradients \citep{teague2018,rosotti2020,garg2022,izquierdo2022,izquierdo2023,Stadler_exoALMA}, self-gravity of the gas \citep{veronesi2021,lodato2023,martire2024,speedie2024,Longarini_exoALMA}, winds \citep{galloway-sprietsma2023} and meridional flows \citep{teague2019a} often fix geometric parameters such as inclination and position angle, as well as the emitting layer height and temperature distributions, even though these quantities themselves are uncertain.
This issue is addressed simply either with Monte Carlo error propagation \citep{andrews2024} or bootstrapping \citep{Galloway_exoALMA}, provided some reasonable quantification of the uncertainty on those parameters.

In this paper we present a first step towards building a method that can in principle address \emph{all} of the above concerns.
We provide a flexible, non-parametric model for intensity distribution in each channel of a spectral cube, including the convolution with the beam in the data generation process, as well as the correlations created between neighbouring pixels.
Samples from the posterior may be generated cheaply due to the analytic posterior, and fed through any cube analysis pipeline to calculate uncertainties and correlations in measurements.
We demonstrate a simple example use case of the model in the generation of probabilistic peak intensity, peak velocity, and line width maps that are unbiased by beam smearing.
We also combine our approach with \textsc{discminer} \citep{izquierdo2021,izquierdo2022,izquierdo2023} to investigate the effect of the more realistic noise model, including spatial correlations, on the precision of the inferred disk model parameters.

The paper is organised as follows: We present the model and analysis methods in Section~\ref{sec:methods}, and we benchmark the method in Section~\ref{sec:benchmark}.
Section~\ref{sec:results} presents our application of the model to a subset of the exoALMA sample, while Section~\ref{sec:discussion} discusses our main findings.
We conclude in Section~\ref{sec:conclusion}.

\begin{figure*}
    \centering
    \includegraphics[width=0.96\textwidth]{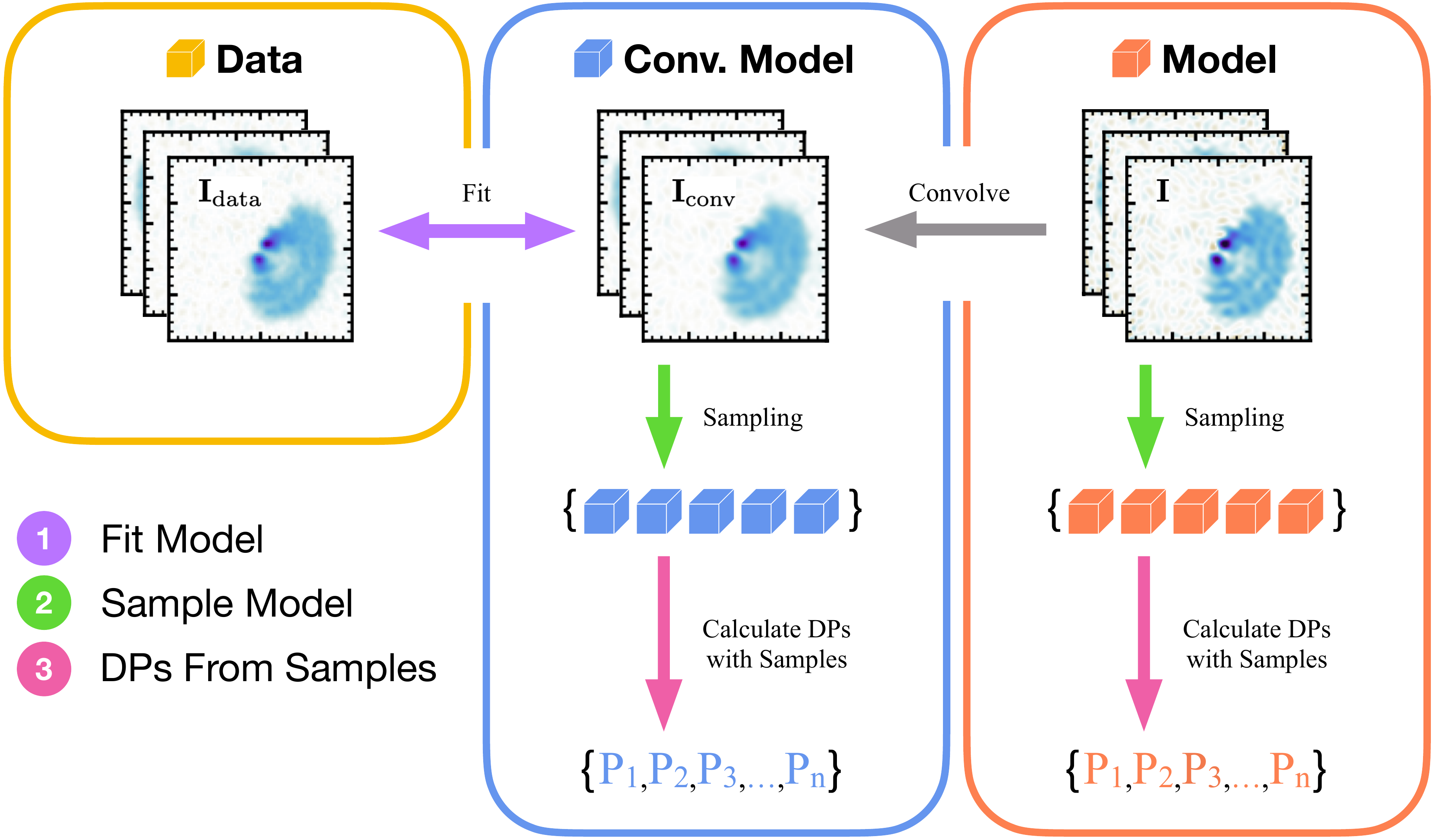}
    \caption{Schematic overview of the methodology for generating probabilistic data products using the non-parametric cube model. The cube graphic represents data cubes, with the orange and blue cubes illustrating posterior model cube samples before and after beam convolution, respectively. The ${\rm P}_i$ values indicate sample data products (DPs) calculated from the corresponding cube samples, colour-coded accordingly. The numbered steps in the bottom-left correspond directly to those in Section~\ref{sec:overview}.}
    \label{fig:overview}
\end{figure*}

\section{Methods} \label{sec:methods}

\subsection{Data} \label{sec:data}

We perform all analysis using the fiducial $^{12}$CO $J=3-2$ cubes \citep{Teague_exoALMA} for HD~135344B, MWC~758, CQ~Tau and J1842.
These have a circular synthesised beam with a full width half maximum (FWHM) of $0\farcs15$, and a channel spacing of $100 \, {\rm ms^{-1}}$.

HD~135344B, MWC~758 and CQ~Tau are used to demonstrate the generation of probabilistic peak intensity, peak velocity and line width maps using our method.
These sources were chosen due to their low inclination, allowing us to treat the line profile as single-peaked due to their lack of backside contamination, and because they all feature large spiral features in CO \citep{casassus2021,boehler2018,wolfer2021,Izquierdo_exoALMA}.
We emphasise that the method does not ultimately require any particular line profile shape, but that this choice allows us to focus on testing downstream data products in a relatively simple setting.

J1842 is used for the case study combining our method with \textsc{discminer}, and was chosen as a representative typical source from the sample.
By this, we mean that it is mid-inclination ($39^\circ$){\bf ,} orbits an approximately $1 \, {\rm M_\odot}$ star \citep{Teague_exoALMA}, and that the channels feature only modest deviations from smooth rotation \citep{Pinte_exoALMA}.
The $^{12}$CO is also only $5\farcs$ across, which aids in computational cost.

We downsample each cube by a factor of 2 in each spatial axis, replacing $2 \times 2$ blocks of pixels with their mean value.
The cubes are also cropped spatially to include only the region where the signal is contained.
Both these steps are done to reduce computational expense, which is discussed further in Section~\ref{sec:discussion}.

\subsection{Non-parametric model} \label{sec:linear_algebra}

\begin{figure*}
    \centering
    \includegraphics[width=0.96\textwidth]{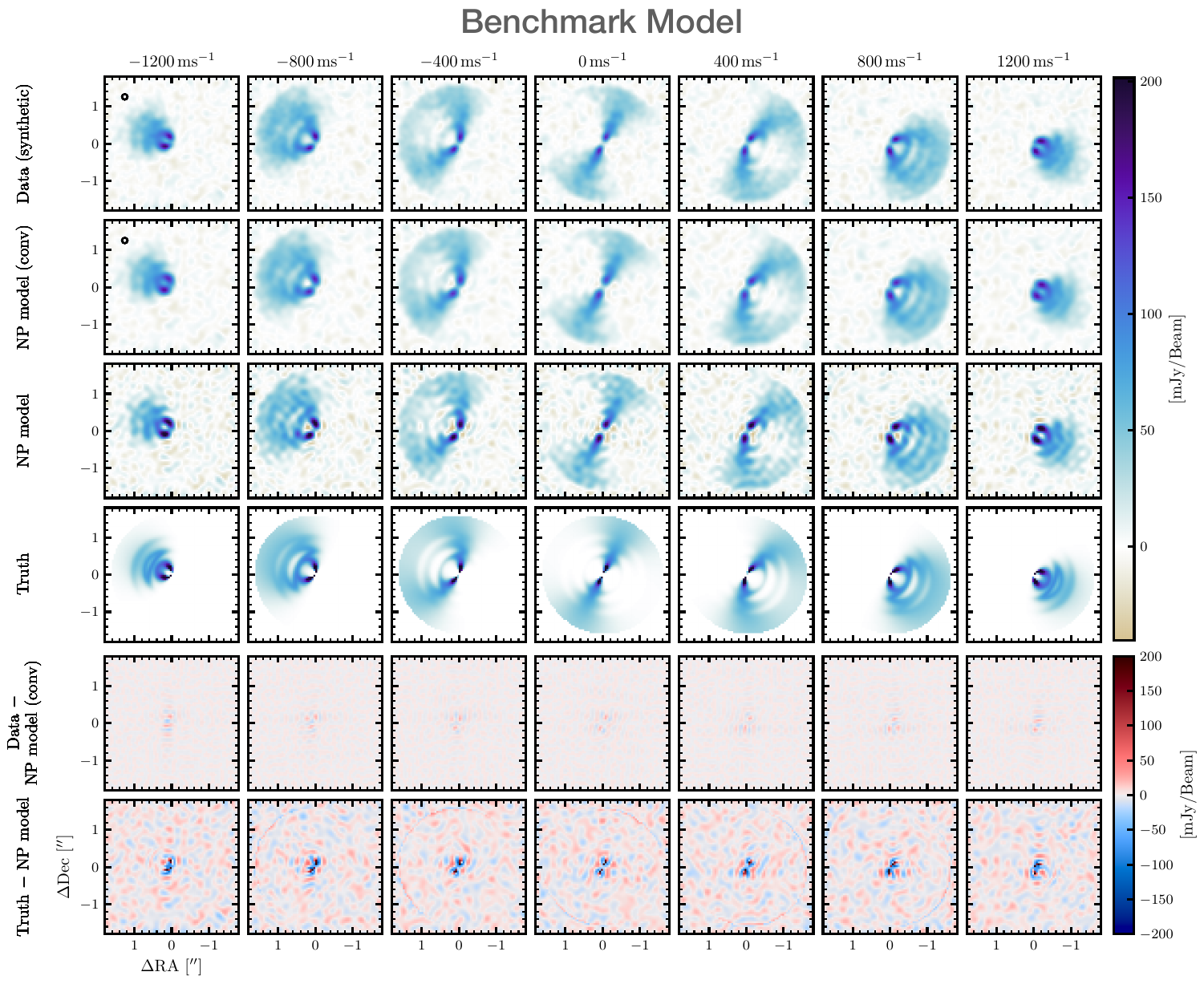}
    \caption{Best-fitting non-parametric (NP) model cube fit to the \textit{benchmark model} synthetic data which has 0\farcs{15} beam and noise comparable to the data. Top row shows synthetic data. Second row shows best fit after beam convolution (beam plotted in top left). Third row shows best fit without beam convolution. Fourth row shows the ground truth for comparison. We see that the model can resolve intensity gradients smeared by the beam. Fifth row shows the residual between the synthetic data and the convolved model. Sixth row shows residuals between model and ground truth. The residuals show the model reconstructs the true intensity distribution up to the noise, except for in the central two beams or so.}
    \label{fig:bench_channels}
\end{figure*}

We elect to directly model the intensity distribution in each channel.
While the model for each channel is independent of the others, for the sake of simplicity we will refer to the collection of model channels comprising the whole cube as a ``model cube''.
The model is non-parametric and consists of a linear summation of orthogonal basis functions, where the coefficients or \emph{weights} of these functions are determined by fitting the model to the data.
This \emph{linear} approach is crucial for allowing for tractable sampling from the model posterior distribution, even with of order $10^4$ model parameters, as the posterior takes the form of a multivariate normal distribution from which we can sample in a few minutes.

Let ${\bf I}$ be a vector containing the model intensities $I_i$ from a particular channel, where $i$ is an index over the $n = n_x n_y$ pixels in the channel.
We model $I_i$ as a weighted sum of $p$ two-dimensional basis functions
\begin{align}
    I(x_i, y_i) &= \sum_{j=1}^p \, X_j f_j \left( x_i, y_i \right) \label{eq:sum_form},
\end{align}
where $x_i$ and $y_i$ are spatial coordinates for pixel $i$, and $X_j$ is the weight associated with basis function $f_j$.
We choose to use a Fourier series for the basis, which consists of orthogonal trigonometric functions in both the $x$ and $y$ directions.
We expand on this choice later in this section.
This allows us to decompose $f_j$ into two identical sets of basis functions, one for each spatial direction, giving
\begin{align}
    I(x_i, y_i) &= \sum_{l=1}^{p_x} \sum_{m=1}^{p_y} \, X_{lm} \left[ g_l \left( x_i \right) \cdot g_m \left( y_i \right) \right] \label{eq:2sum_form},
\end{align}
where $p = p_x p_y$, $l = j \, {\rm mod} \, p_x$, $m = \lfloor j / p_x \rfloor$, $f_j(x,y) = g_l(x) \cdot g_m(y)$, and $\lfloor \cdot \rfloor$ denotes the floor function.
The basis functions $g_l$ are defined as (for both $l$ and $m$)
\begin{align}
    g_l (x) = \begin{cases}
    \cos{(\omega_l x)} \quad {\rm for} \, l \, {\rm odd} \\
    \sin{(\omega_l x)} \quad {\rm for} \, l \, {\rm even}
    \end{cases}, \quad \omega_l = \frac{\pi}{L} \left\lfloor \frac{l}{2} \right\rfloor,
\end{align}
where $L$ is a length scale in the $x$ space.
We choose spatial coordinates $x_i,y_i$ for the pixels uniformly spaced between 0 and 1, and set $L=3$ following \citet{hoggvillar}.
By defining the entries of $n_x \times p_x$ matrix ${\bf A}_x$ as
\begin{align}
    \left[ {\bf A}_x \right]_{ij} = g_j \left(x_i\right),
\end{align}
and similarly for ${\bf A}_y$, we can rewrite Eq.~\eqref{eq:2sum_form} simply as
\begin{align}
    {\bf I} = \left({\bf A}_{x} \otimes {\bf A}_{y}\right) {\bf X} = {\bf A}_{xy} {\bf X} \label{eq:I_from_X}
\end{align}
where ${\bf X}$ is a vector of length $p$ containing the Fourier weights $X_j$, ${\bf A}_{xy}$ is an $n\times p$ matrix, and $\otimes$ is the Kronecker product.
We note that our choice of basis functions does not constrain the model intensities to be positive.
While the physical intensity is strictly positive, the intensity in the synthesised images is not.
Since we only have access to noisy measurements rather than the true underlying intensities, allowing the model to take non-positive values is a deliberate and justified choice.

Let $h(x,y)$ be a function representing the kernel of the beam, a circular 2D gaussian with FWHM~$=0\farcs{15}$ for our purposes, but it may be of any form in general.
Defining the entries of $n \times n$ matrix ${\bf H}$ as
\begin{align}
    \left[{\bf H}\right]_{ij} = h(x_i - x_j, y_i - y_j), \label{eq:conv_matrix}
\end{align}
we can find the vector of the beam-convolved intensities ${\bf I}_{\rm conv}$ from
\begin{align}
    {\bf I}_{\rm conv} = {\bf H} {\bf I} = {\bf H} {\bf A}_{xy} {\bf X} = {\bf A} {\bf X}, \label{eq:Iconv_from_X}
\end{align}
where the last equality is from defining the \emph{design matrix} ${\bf A} = {\bf H} {\bf A}_{xy}$.

Defining ${\bf I}_{\rm data}$ as a vector containing the actual data observed from a particular channel, we can write our overall model as
\begin{align}
    {\bf I}_{\rm data} = {\bf A} {\bf X} + \boldsymbol{\epsilon},
\end{align}
where $\boldsymbol{\epsilon}$ is a random vector of length $n$ representing the noise in the image.
We will assume that the noise follows a multivariate normal distribution (hereafter just normal distribution)
\begin{align}
    \boldsymbol{\epsilon} \sim \mathcal{N} \left( {\bf 0}, {\bf C} \right),
\end{align}
where the $\sim$ notation denotes that $\boldsymbol{\epsilon}$ is \emph{drawn from} the normal distribution $\mathcal{N}$ with zero mean vector and $n \times n$ covariance matrix ${\bf C}$.
We assume that the noise is spatially correlated on the scale of the beam which gives
\begin{align}
    \left[{\bf C}\right]_{ij} = \sigma_i \sigma_j \left[ {\bf H} \right]_{ij} / {\rm max} \left( {\bf H} \right),
\end{align}
where $\sigma_i$ is the marginal standard deviation of the noise in pixel $i$.
In practice, we assume the same $\sigma$ for all pixels in the cube, which we estimate from the standard deviation of the data in channels with no significant signal.

We emphasise that this approach of modelling the image intensities only, and \emph{not} treating the finite sampling of the $uv$-plane, is a non-trivial simplification.
Ideally we would forward model the visibilities, and include a treatment of the $uv$ sampling function, however we leave this for future work.

The model as described so far constitutes a \emph{linear regression model}, even though the basis functions themselves are non-linear.
The typical approach to fitting the above model would be to perform ordinary least squares (OLS), where one finds the ``best'' parameter vector ${\bf X}$, which we will label $\hat{\bf X}$, by minimising the residuals squared
\begin{align}
    \hat{\bf X} = \argmin_{\bf X} \lvert\lvert {\bf I}_{\rm data} - {\bf A} {\bf X} \rvert\rvert^2,
\end{align}
which may be solved via the matrix pseudoinverse\footnote[1]{The matrix ${\bf A}^{\top} {\bf A}$ is only invertible for $p < n$. We use $p > n$ in this paper, but this is not an issue as long as the regularisation term $\left[\Lambda\right]_{jj}$ we introduce is strictly positive for all $j$. This is the case, see Eq.~\eqref{eq:lambda_jj} and Eq.~\eqref{eq:w_func}.} \citep{moore1920,penrose1956}
\begin{align}
    \hat{\bf X} = \left( {\bf A}^\top {\bf A} \right)^{-1} {\bf A}^\top {\bf I}_{\rm data}.
\end{align}
This approach may be extended to include the covariance matrix, and can be understood as a simple reweighting of the data by the inverse of their covariances
\begin{align}
    \hat{\bf X} = \left( {\bf A}^\top {\bf C}^{-1} {\bf A} \right)^{-1} {\bf A}^\top {\bf C}^{-1} {\bf I}_{\rm data}, \label{eq:GLS}
\end{align}
which is called the generalised least squares (GLS) estimate.
The GLS estimate has some desirable properties, namely that it is the best linear unbiased estimator\footnote[2]{This is because GLS is equivalent to OLS under an affine transformation and OLS is the best linear unbiased estimator via the Gauss-Markov theorem \citep[e.g.][]{kutner2004}}, and that it maximises the model likelihood.

This approach will always result in a unique solution for $\hat{\bf X}$ and the reconstructed ${\bf I}_{\rm conv}$.
However, we are interested in the model intensities prior to beam convolution ${\bf I}$ given by Eq.~\eqref{eq:I_from_X}, which becomes increasingly poorly behaved as the number of basis functions $p$ is increased.
Because ${\bf H}$ acts in the design matrix to dampen signal contributed from high frequency Fourier modes, there is little information available to constrain the weights of those modes.
${\bf I}$ therefore has many possible values that result in similar ${\bf I}_{\rm conv}$.
This problem is identical to the linear deconvolution problem in the presence of noise, which has been studied extensively for at least 80 years \citep[e.g.][]{wiener1949,oldenburg1981}.

A natural solution is to provide additional contraints on $\hat{\bf X}$ that will result in an ${\bf I}$ that we believe to be more sensible \cite[e.g.][]{gelman2013}.
From a frequentist perspective this additional constraint is called a regulariser, while the (equivalent) Bayesian view is that of a prior probability distribution over possible ${\bf X}$.
Here we will mostly adopt the language of the latter as we wish to use the posterior probability distribution (the resultant distribution from updating the prior with the data via Bayes' theorem) in our analysis.
We first assume some generic normal prior on ${\bf X}$
\begin{align}
    {\bf X} \sim \mathcal{N} \left( \boldsymbol{\mu}, \boldsymbol{\Lambda} \right),
\end{align}
where $\boldsymbol{\mu}$ is a mean vector of length $p$ and $\boldsymbol{\Lambda}$ is a $p \times p$ covariance matrix.
The posterior distribution over ${\bf X}$ has an analytic form thanks to the linearity of the model, and because both the noise and prior are normal distributions
\begin{align}
    p \left( {\bf X} \,|\, {\bf I}_{\rm data} \right) = \mathcal{N} \left( {\bf X} \,|\, \hat{\bf X}, \boldsymbol{\Sigma} \right), \label{eq:posterior}
\end{align}
where the right-hand side is shorthand notation for the probability density of the normal distribution with mean vector $\hat{\bf X}$ and covariance matrix $\boldsymbol{\Sigma}$ evaluated at some ${\bf X}$.
Additionally \citep[e.g.][]{hogg2020}
\begin{align}
    \hat{\bf X} &= \boldsymbol{\Sigma} \left( {\bf A}^{\top} {\bf C}^{-1} {\bf I}_{\rm data} + \boldsymbol{\Lambda}^{-1} \boldsymbol{\mu}\right) \label{eq:Xhat} \\
    \boldsymbol{\Sigma} &= \left( {\bf A}^\top {\bf C}^{-1} {\bf A} + \boldsymbol{\Lambda}^{-1}\right)^{-1} \label{eq:Sigma},
\end{align}
where the differences between the above and Eq.~\eqref{eq:GLS} can be understood as reweighting the predictions using the prior.


We choose a prior that penalises high frequency basis functions more than low frequency ones, in order to encourage smoothness.
This can be achieved with feature weighting \citep{rauhut2016, bah2016, xie2022}, where we set $\boldsymbol{\mu} = {\bf 0}$ and the prior variance for each basis function using some weighting function $w$ that decreases with frequency
\begin{align}
    \left[ \boldsymbol{\Lambda} \right]_{jj} = \frac{1}{\lambda} \left[ w \left( \omega_l, \omega_m \right)\right]^2, \label{eq:lambda_jj}
\end{align}
recalling that $\omega_l$ and $\omega_m$ are the $x$ and $y$ direction frequencies for basis function $f_j$, and we choose \citep{hoggvillar}
\begin{align}
    w(\omega_l, \omega_m) = \left( \frac{\pi}{2} \right)^{1/4} \left( \frac{1}{s} + s \left[\omega_l^2 + \omega_m^2\right] \right)^{-1/2}, \label{eq:w_func}
\end{align}
where both $\lambda$ and $s$ are \emph{hyperparameters} for our model, that encode how strong our prior is.
$\lambda$ is an overall scale that uniformly scales the variance across all $X_j$, while $s$ controls how quickly the variance decreases with the frequency of the basis.
We somewhat arbitrarily choose values for the hyperparameters of $\lambda=s=0.1$, since we found them to be performant in terms of predicting the data (see Section~\ref{sec:benchmark}) without fine-tuning.
We discuss the caveats associated with fixing the hyperparameters to arbitrarily values, in particular the impacts on the resultant model uncertainties, in Appendix~\ref{sec:hyperparam}.

We also choose the \emph{over-parameterised} case, where we have more model parameters than data points $p>n$.
We set the number of basis functions in each direction $p_x, p_y$ to be 1.25 times the number of pixels in each direction $n_x, n_y$, rounded to the nearest integer.
We justify this choice by returning to our choice of basis.
It was realised in the last few years that a Fourier basis is an effective way to represent stationary kernels for Gaussian process (GP) regression \citep{hensman2018,hoggvillar, greengard2022}.
Gaussian processes are distributions over functions \citep[e.g.][]{rasmussen2006}.
In the limit $p \rightarrow \infty$, our basis and prior choice is equivalent to a GP prior on the model intensities
\begin{align}
    I({\bf x}) \sim \mathcal{GP} \left( 0, k\left( {\bf x}, {\bf x}' \right)\right)
\end{align}
where ${\bf x} = \left[x, y\right]^\top$, and $k$ is a function encoding the model covariance between two points, usually called the kernel.
Specifically the equivalence is given by
\begin{align}
    \lim_{p \rightarrow \infty} \sum_{j=1}^p \left[\boldsymbol{\Lambda}\right]_{jj} f_j \left( {\bf x} \right) f_j \left( {\bf x}' \right) = k \left( {\bf x}, {\bf x}' \right),
\end{align}
and we refer the reader to \citet{hoggvillar} and \citet{hensman2018} for further detail.
\citet{tobar2023} recently showed that GP priors can be used in this manner to solve the noisy deconvolution problem in a Bayesian setting, outperforming conventional approaches such as the Wiener filter while simultaneously providing uncertainties on the recovered signal.
While our approach is only an approximation of theirs, kernels with rapidly decaying Fourier transforms are well approximated by an equispaced Fourier series for even relatively small $p$ \citep{greengard2022}.
Our choice of weighting function given in Eq~\eqref{eq:w_func} is equivalent to the exponential kernel \citep{rasmussen2006,hoggvillar}, and its Fourier transform decays quadratically.

\subsection{Fitting and sampling} \label{sec:fitting}

We perform the fits by calculating the posterior mean ${\bf \hat{X}}$ and covariances $\boldsymbol{\Sigma}$ using Eq.~\eqref{eq:Xhat} and \eqref{eq:Sigma} respectively.
Note that while the ${\bf \hat{X}}$ must be calculated for each channel separately, $\boldsymbol{\Sigma}$ depends only on quantities which are identical for all channels and only needs to be calculated once for each cube.
The best-fitting model channels are then generated as ${\bf I} = {\bf A}_{xy} \hat{\bf X}$, while the best-fitting convolved model is ${\bf I}_{\rm conv} = {\bf A} \hat{\bf X}$.
We additionally draw 100 posterior samples of ${\bf X}$ for each channel from the posterior $\mathcal{N} ( \hat{\bf X}, \boldsymbol{\Sigma} )$, and convert them to posterior samples of {\bf I} and ${\bf I}_{\rm conv}$ in the same way as for the best fit.
``Sample cubes'' are assembled by grouping a unique sample of each channel together, yielding 100 model cube samples, both before and after beam convolution.
Details about the numerical methods and accuracy considerations for both the fitting and sampling are given in Appendix~\ref{sec:numerical}.

\subsection{General probabilistic data products} \label{sec:overview}

\begin{figure*}
    \centering
    \includegraphics[width=0.9\textwidth]{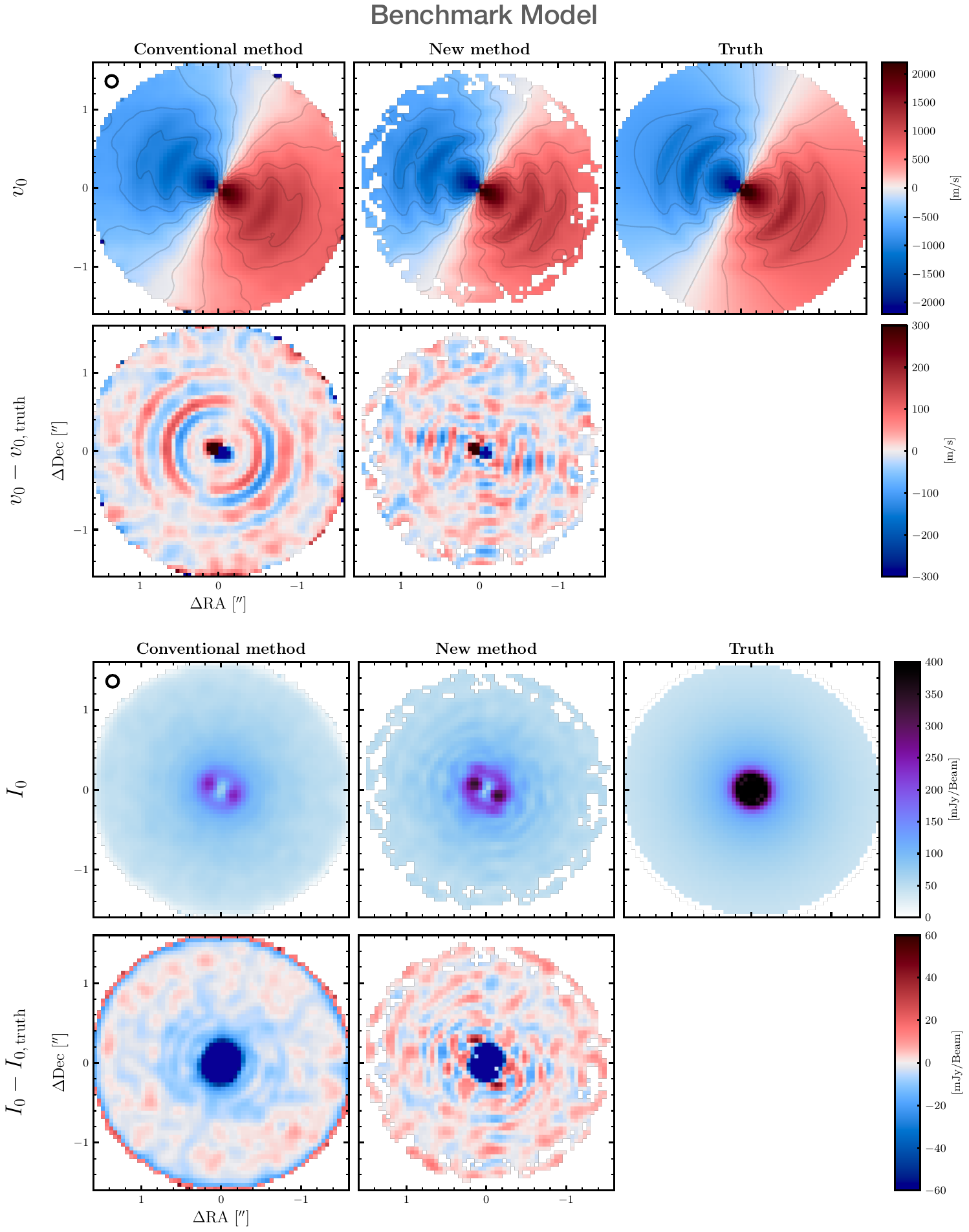}
    \caption{Top: Peak velocity maps $v_0$ extracted from the \textit{benchmark model} synthetic data, using both the conventional method (left; see Section~\ref{sec:methods_moments}) and our method (right). Right-most panel shows the true \textit{benchmark model} line-of-sight velocities $v_{0, \, {\rm truth}}$. $v_0$ maps show also iso-velocity contours spaced by $200 \, {\rm m s^{-1}}$, and the beam is shown in the top-left panel. Second row shows the difference between the recovered $v_0$ maps and the truth. Our method debiases the $v_0$ map and recovers the full amplitude of the substructures. Bottom: Same as above except showing recovered peak intensity maps $I_0$.}
    \label{fig:vI_hd135_model}
\end{figure*}

Figure~\ref{fig:overview} illustrates the three steps involved in using applying our method generally to generate probabilistic data products from spectral cube data, which are as follows:
\begin{enumerate}
    \item The model is fit to the data cube to determine the posterior distribution for each channel. 
    \item Posterior model cube samples are generated. 
    \item The data products of interest are calculated independently from each cube sample, yielding posterior samples of each product.
\end{enumerate}
As shown in Figure~\ref{fig:overview} and indicated in Section~\ref{sec:fitting}, the posterior model cube samples can be generated either beam-convolved (blue in Figure~\ref{fig:overview}) or not (orange).
This choice would generally depend on the specific interests of the investigator.
In our analysis, we will use the unconvolved cube samples for the peak intensity, peak velocity and line width maps.
For the case study using \textsc{discminer}, we use the convolved cube samples since \textsc{discminer} forward models the effect of the beam.

The posterior data product samples can either be used to reconstruct the marginal posterior distribution over that product, or just to find a most likely value with associated uncertainties.
The distribution is estimated simply from a histogram or kernel density estimate, while the best value and uncertainties can be taken from the median or mean, and chosen quantiles respectively \citep[e.g.][]{hogg2018}.

\subsection{Probabilistic moment maps} \label{sec:methods_moments}

As a simple example for a probabilistic data product, we generate peak intensity, peak velocity and line width maps, commonly called moment maps, following our method.
We do this for HD~135344B, CQ~Tau, and MWC~758 as outlined in Section~\ref{sec:data}.

We fit Gaussian line profiles to each pixel of each sample cube 
\begin{align}
    f(v_{\rm ch}, {\boldsymbol{\theta}}_{0}) = I_0 \exp{\left(- \frac{1}{2} \left[ \frac{v_{{\rm ch}} - v_0}{w_0} \right]^2 \right)}, \label{eq:gaussian_line}
\end{align}
where ${\boldsymbol{\theta}}_{0} = \left[I_0, v_0, w_0 \right]$, $I_0$ is the peak intensity, $v_0$ is the peak or centroid velocity, and $w_0$ is the line width.
Each line profile in each cube sample will have it's own best-fitting ${\boldsymbol{\theta}}_{0}$, but we neglect indices over these in the notation for simplicity.
We use maximal likelihood estimation to perform the fit, in practice minimising the negative log likelihood given by
\begin{align}
    \log \mathcal{L}_m \left( {\bf I} \, | \, {\boldsymbol{\theta}}_{0} \right) = - \frac{1}{2\sigma^2} \sum_{i}^{n_{\rm ch}} \left[ I_i - f(v_{{\rm ch},i}, {\boldsymbol{\theta}}_{0}) \right]^2,
\end{align}
where $i$ is an index over channels, and we assume the noise $\sigma$ is normally distributed.
We use the L--BFGS--B optimisation algorithm \citep{broyden1970,fletcher1970,goldfarb1970,shanno1970,liu1989} implemented in SciPy's \codesnip{optimize.minimize} function \citep{virtanen2020} to fit.
A noise cut-off of three times the cube RMS was applied to avoid fitting lines with low signal-to-noise.

After applying the above procedure to each sample cube (pre-convolution), we can estimate best fits and associated uncertainties over $I_0, v_0, w_0$ for each data cube, as described in Section~\ref{sec:overview} above.
The best-fitting ${\boldsymbol{\theta}}_{0}$ for each line is taken as the median across the 100 cube samples, while the $1\sigma$ uncertainty is estimated from the standard deviation.
We present these results in Section~\ref{sec:results_moments}.

We also fit Gaussian line profiles to the data directly, as is typically done in studies of kinematics or other disk properties, for comparison.
In our results we refer to the probabilistic maps described above as the ``new method'', and the fit directly to the data as the ``conventional method''.
We emphasise that the important distinction is the use of the non-parametric model and sampling in our method, over the direct fit to the data in the conventional method.

\subsection{Discminer case study} \label{sec:methods_discminer}

\textsc{Discminer} is a statistical inference tool for protoplanetary disk properties that uses a forward model of line emission to constrain global properties of the disk including the mass of the central star, the inclination and position angle, and the radial height and intensity profiles \citep{izquierdo2021}.
\textsc{Discminer} is used extensively in the analysis performed by exoALMA \citep{Izquierdo_exoALMA}, as the parametric disk model it provides is ideal for quantifying localised deviations in for example the rotation curve \citep{Stadler_exoALMA,Longarini_exoALMA} or pressure scale height \citep{Stadler_exoALMA}.

\textsc{Discminer} uses Bayesian inference to estimate the posterior distribution of the disk model, allowing for quantification of the model uncertainties.
However, it uses a simplified noise model where the correlations between nearby pixels in the cube are neglected.
This simplification is to avoid the increased computational expense incurred evaluating a multivariate normal likelihood during the Markov Chain Monte Carlo (MCMC) sampling.
Such correlations are known to widen the posterior distribution and so their exclusion will in general result in underestimated uncertainties \citep[e.g.][]{gelman2013}, and predicting the factor by which they are underestimated is not straightforward.

Our non-parametric model of the channels can cheaply incorporate spatial correlations in the noise, because it is linear (see Section~\ref{sec:linear_algebra}).
This allows us to combine our approach with \textsc{Discminer} to assess the impact of these correlations on inference precision.
We use J1842 as a test case for this combined approach.

To prevent ambiguity, we will refer to posterior samples from our non-parametric model as ``cube samples'', and posterior samples from the \textsc{Discminer} model as ``\textsc{Discminer} samples''.
We generated 30 cube samples according to the procedure outlines in Section~\ref{sec:overview}, in this case using the \emph{convolved} samples as \text{Discminer} forward models the effect of the beam.
We fit each cube sample with \textsc{Discminer}, obtaining many posterior distributions, where each is conditioned on a single cube sample.
Finally, we estimate the combined posterior by aggregating the posterior \textsc{Discminer} samples from all fits.

Additionally, we also re-fit \textsc{Discminer} to the fiducial J1842 data cube downsampled by the same factor as used for the non-parametric model fit (see Section~\ref{sec:data}).
The production \textsc{Discminer} fit for J1842 used across the other exoALMA papers \citep{Teague_exoALMA,Izquierdo_exoALMA} used downsampling of $6\times6$ pixel blocks, whereas we used $2\times2$.
We compare these results with the combined approach in section~\ref{sec:results_discminer}.

\section{Benchmark} \label{sec:benchmark}


\subsection{Synthetic data and fit} \label{sec:bench_data}

For the benchmark model, we assume the line in each pixel is Gaussian, and choose radial intensity, velocity, and line width profiles, as well as other model parameters to create similar looking synthetic data to HD~135344B.
The specific radial profiles we choose are given in Appendix~\ref{sec:bench_model}.
For the model we use central mass $M_\star = 1.61 \, {\rm M_\odot}$, inclination $i = 16.11^\circ$ and position angle ${\rm PA} = 242.94^\circ$ \citep{Izquierdo_exoALMA}, and a distance of $135 \, {\rm pc}$ \citep{gaiacollaboration2023}.
The true velocity, intensity and line width maps $v_{0, \rm truth}$, $I_{0, \rm truth}$ and $w_{0, \rm truth}$ are calculated from the radial profiles by rotating sky coordinates to match the inclination and position angle of the model, and for the velocities by taking only the line-of-sight component.
These maps are shown in the upper-right panels of Figures~\ref{fig:vI_hd135_model} and \ref{fig:w_hd135_model}.

True channels are created using the true maps using the Gaussian line profile in Eq.~\eqref{eq:gaussian_line}, with $v_{\rm ch}$ in increments of $100 \, {\rm ms^{-1}}$.
These channels are shown in the 4th row of Figure~\ref{fig:bench_channels}.
We then convolved the channels spatially with a circular Gaussian beam with FWHM~$=0\farcs{15}$ to match the data, and added beam-correlated noise with scale $\sigma = 20 \, {\rm mJy}$.
This synthetic data is shown in the top row of Figure~\ref{fig:bench_channels}.

We fit the synthetic data using the non-parametric model.
The best-fitting model and convolved model are shown in rows 3 and 4 of Figure~\ref{fig:bench_channels}.
The residuals between the synthetic data and the fit (5th row) are all on the scale of the noise or less, with the most significant difference found near the centre of the disk.
By comparing the unconvolved model fit with the true channels, and with the data, we see that the model is capable of recovering the steep intensity gradients co-located with the substructures that are partially smeared out by the beam.
Similarly, the intensity of the channels near the centre is better resolved in the model compared to the data, although the model is not able to reconstruct the very sharp edges present in the true channels in this region.
The latter effect is likely explained by the model prior, which encourages smoothness.
The last row of the figure shows the residuals between the model and true channels, which are mostly just noise except in the inner regions.
This indicates that the model effectively provides an unbiased estimate of the intensity, except within the central few beams.

Note also that the model \emph{contains noise}.
Due to its non-parametric nature, the model effectively does not distinguish between signal and noise in the data.
Indeed, the model before beam convolution is noiser than the data itself due to the uncertainty associated with the information lost through beam convolution.

\subsection{Moment maps and substructures} \label{sec:bench_moments}

\begin{figure*}
    \centering
    \includegraphics[width=0.9\textwidth]{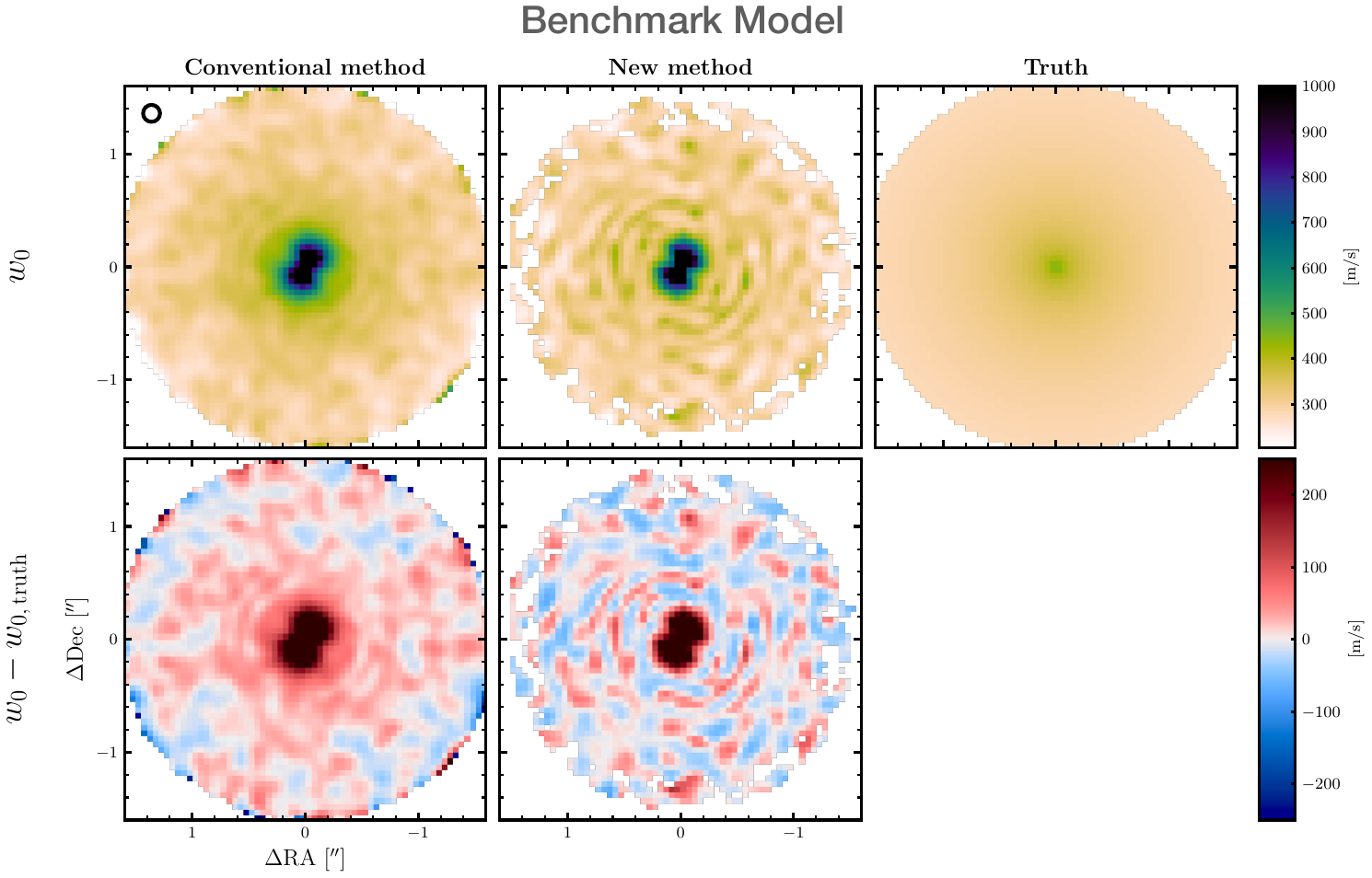}
    \caption{The same as Figure~\ref{fig:vI_hd135_model} except showing recovered line width maps $w_0$.}
    \label{fig:w_hd135_model}
\end{figure*}

The middle columns of Figures~\ref{fig:vI_hd135_model} and \ref{fig:w_hd135_model} present the probabilistic $v_0$, $I_0$ and $w_0$ maps calculated following our new method.
These may be compared to maps calculated with the conventional method (see Section~\ref{sec:methods_moments}), shown in the left columns.
The differences between each method and the true map used for the synthetic data are shown in the bottom rows.

Comparing the peak velocity maps (Figure~\ref{fig:vI_hd135_model} top plot), we can see from the bottom panels that the new method provides a less biased estimate of the velocity along the ring substructures. 
As expected, the conventional method is biased by beam smearing, resulting in the underestimation of the velocity perturbations of 50--200$ \, {\rm ms^{-1}}$.
This bias is not present in our new method.
This effect can also be seen in the contours drawn in the top panels, which are commonly used to trace disk substructures \citep[e.g.][]{calcino2022}.
The ``wiggles'' in the contours of the conventional method are smaller than in the true map, while those in the new method map more accurately match the truth.
Both methods underestimate the velocities in the central beam or so, although the new method performs better as the biased region is smaller.
The new method contains some artefacts aligned with the vertical and horizontal directions, which could be due the 2D Fourier basis causing a preferred direction in the model.

The peak intensity maps (Figure~\ref{fig:vI_hd135_model} bottom plot) show that the conventional map systematically underestimates the peak intensity in most of the disk, although only by around 10\%.
The new method does not contain this bias, throughout most of the disk.
The central few beams are biased low in both methods, except for the two lobes near the high-velocity wings where the new method performs better.
The line width maps (Figure~\ref{fig:w_hd135_model}) are similar to the intensity except that they are biased high in the conventional method and the bias is greater, around 25\%.
The new method is again unbiased apart for the same inner region.

\section{Results} \label{sec:results}

\begin{figure*}
    \centering
    \includegraphics[width=0.96\textwidth]{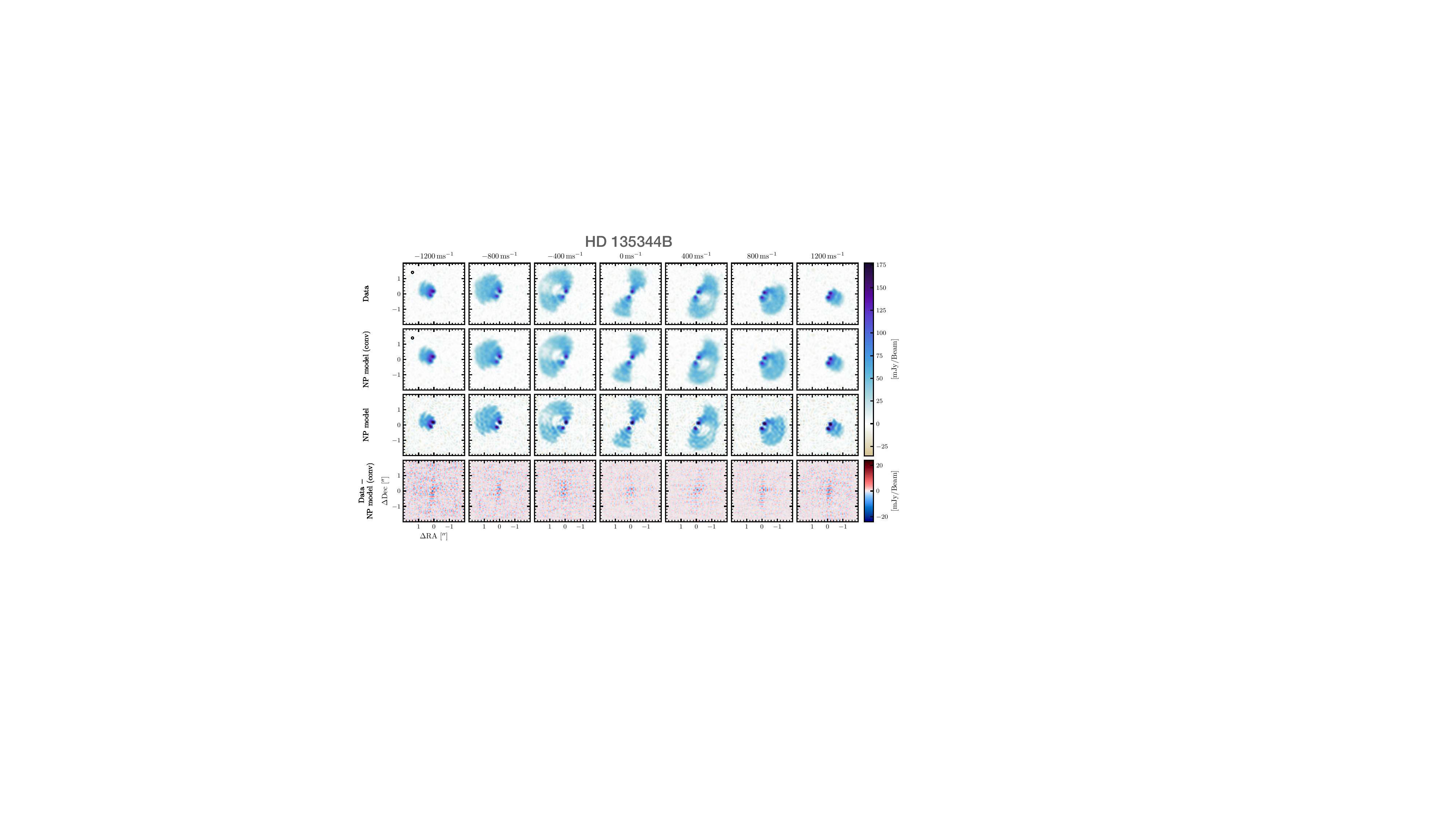}
    \caption{Best-fitting non-parametric (NP) model cube for HD 135344B. Top row shows the data. Second row shows the best fit after beam convolution. Third row shows the best fit prior to beam convolution. Fourth row shows the residual between the data and convolved model.}
    \label{fig:channels}
\end{figure*}

Figure~\ref{fig:channels} shows the best-fitting non-parametric model for HD~135344B compared with the data.
As in the benchmark, the residuals are all on the scale of the noise or less, and again the most significant difference is found towards the centre.
The model prior to convolution contains steeper intensity gradients near the obvious arc-like substructures found in the channels, although the noise is increased.

\subsection{Moment maps} \label{sec:results_moments}

\begin{figure*}
    \centering
    \includegraphics[width=0.92\textwidth]{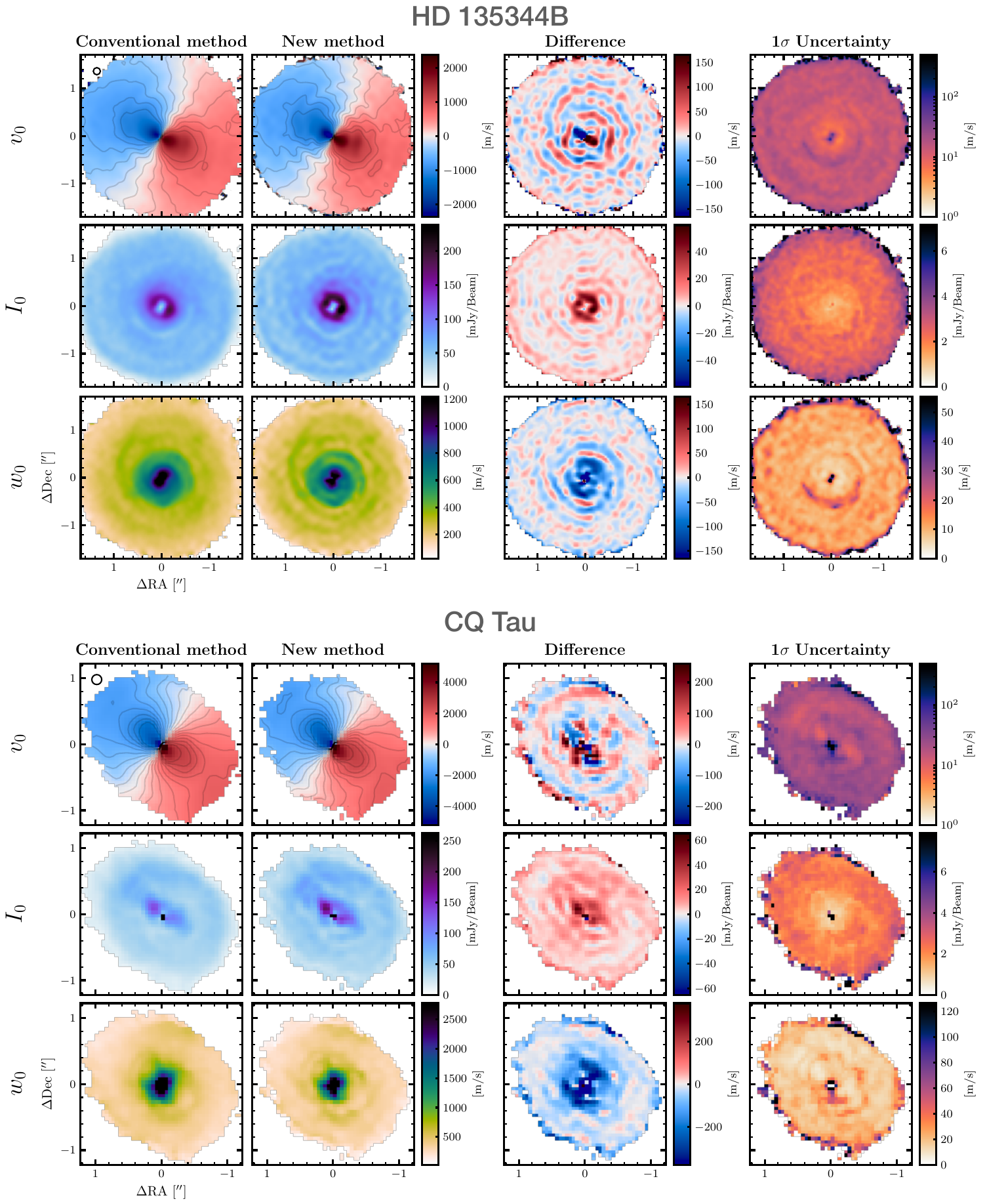}
    \caption{Peak velocity, peak intensity, and line width maps for HD~135344B (top) and CQ~Tau (bottom). Left column shows the results using the conventional method. Second column shows the results using our new method. Third column shows the difference of these. Fourth column shows the $1\sigma$ uncertainty on the results from the new method. As in the benchmark, the new method better resolves the velocity perturbations associated with substructures.}
    \label{fig:moments_hd135_cqtau}
\end{figure*}

\begin{figure*}[!t]
    \centering
    \includegraphics[width=0.92\textwidth]{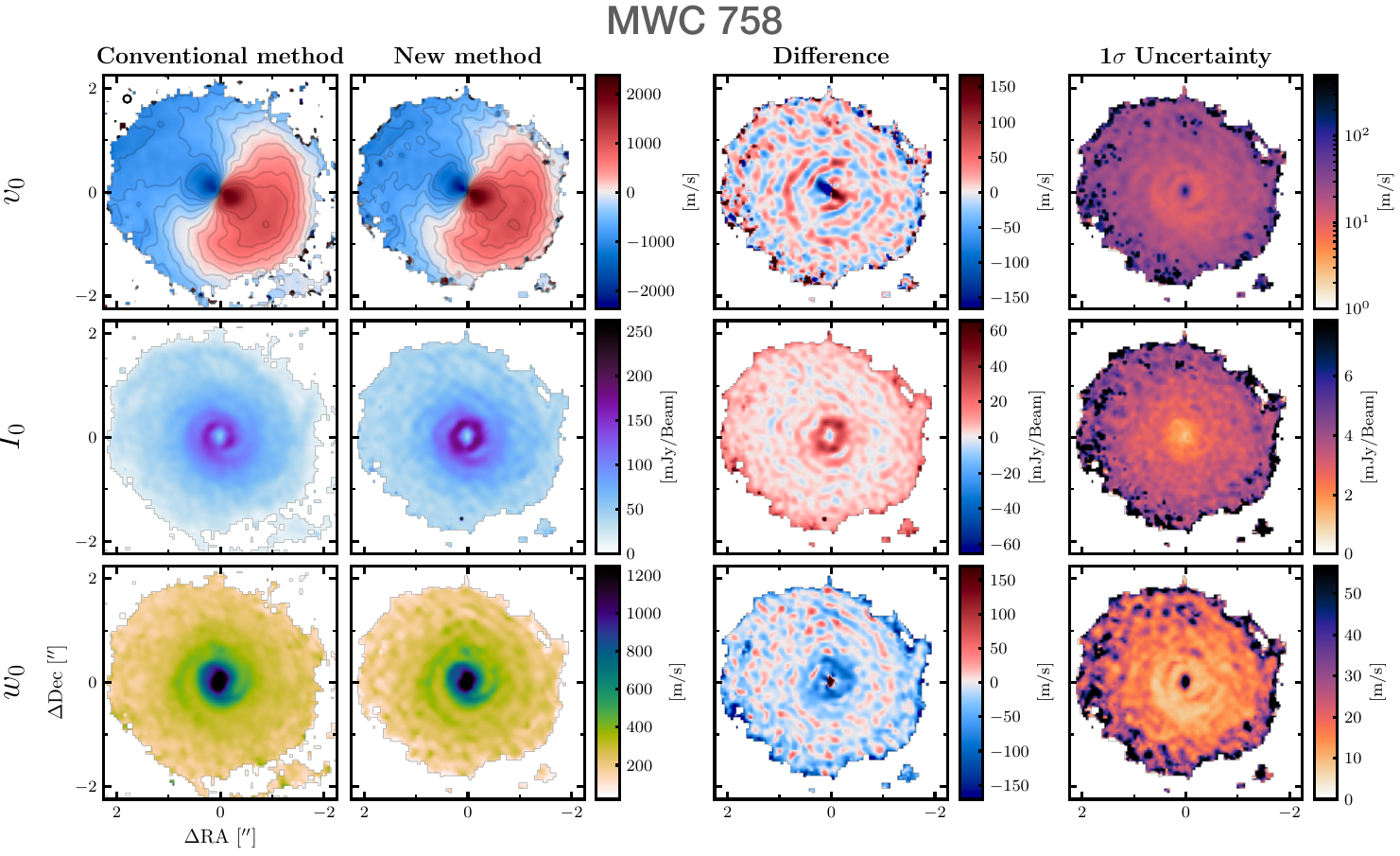}
    \caption{The same as Figure~\ref{fig:moments_hd135_cqtau} except for MWC~758.}
    \label{fig:moments_mwc758}
\end{figure*}

The probabilistic $v_0$, $I_0$ and $w_0$ maps for HD~135344B, CQ~Tau and MWC~758 are presented in Figures~\ref{fig:moments_hd135_cqtau} and \ref{fig:moments_mwc758}, and we compare our method to the conventional approach.
The difference between the two methods is also shown, as well as the $1\sigma$ uncertainty for each of the maps.
There are common features shared across the results from the three sources.
As in the benchmark, the new method finds greater peak intensities and reduced line widths throughout the disk in each case.
The new method also finds greater velocity perturbation amplitudes associated with disk substructures, particularly for HD~135344B.
This difference in recovered perturbation size also leads to coherent spiral structures in the panels showing the difference between the $v_0$ maps for each method.
These disks are all known to contain spirals \citep{casassus2021,boehler2018,wolfer2021,Izquierdo_exoALMA}, and so the model's ability to recover them, in combination  with the benchmark, provides reassurance that it is accurately reflecting the data rather than introducing artificial features.

The maps for HD~135344B (Figure~\ref{fig:moments_hd135_cqtau}, top) show clear signs of large-scale spiral structures spanning across the whole face of the disk in both the velocities and line widths.
Our method resolves both of these better than the conventional one, finding velocity perturbations along the spiral 50--100$\, {\rm ms^{-1}}$ larger.
The plotted contours on the $v_0$ maps (top row) show how this difference manifests as much larger and sharper wiggles in the iso-velocity curves, similar to that in the benchmark.
Interestingly, the line widths (bottom row) \emph{along} the spiral are found to be larger than with the conventional method, even though the general trend is lower.
The combined effect of smaller line widths in most of the disk, and larger line widths along the spiral, results in increased contrast of the spiral in the $w_0$ map.
The $I_0$ map (middle row) on the other hand does not differ much between the methods in terms of substructures, and the difference plot shows a general trend towards higher intensities in the new method.

The new method also recovers stronger velocity perturbations in CQ~Tau (Figure~\ref{fig:moments_hd135_cqtau}, bottom), with a difference along the large one-armed spiral in the top-left of 50--150 ${\rm ms^{-1}}$.
Unlike for HD~135344B, the $w_0$ map (bottom row) does not show significant substructure, but the $I_0$ map (middle row) does show hints of the spiral found in $v_0$ (top row), and is more obvious in the new method result.

MWC~758, unlike the other two sources, shows clear signs of substructure in all 3 maps (Figure~\ref{fig:moments_mwc758}, bottom).
Our method again finds larger velocity perturbations (top row) associated with the spiral structure, in the 50--150 ${\rm ms^{-1}}$ range, and the contour wiggles are sharper.
Both the $w_0$ and $I_0$ maps (middle and bottom rows) show increased contrast of the spiral.

The $1\sigma$ uncertainty on the $v_0$ maps correlate with $I_0$ in all 3 sources, which is to be expected as the line centre measurement precision is ultimately set by the signal-to-noise ratio.
HD~135344B shows the highest precision, with an uncertainty of 5--10$\, {\rm ms^{-1}}$ across most of the disk (note the same colour map scaling for the $v_0$ uncertainties in all figures).
Both CQ~Tau and MWC~758 have uncertainties in the range 10--30$\, {\rm ms^{-1}}$.
This demonstrates that, despite the spatially correlated noise, we can still constrain the line centre with a precision comparable to or better than the native spectral resolution of 26 $\rm{m s^{-1}}$ \citep{Loomis_exoALMA}, which ultimately sets the fundamental velocity resolution in the absence of additional information about the line profile.

\subsection{Discminer} \label{sec:results_discminer}

\begin{figure*}
    \centering
    \includegraphics[width=0.92\textwidth]{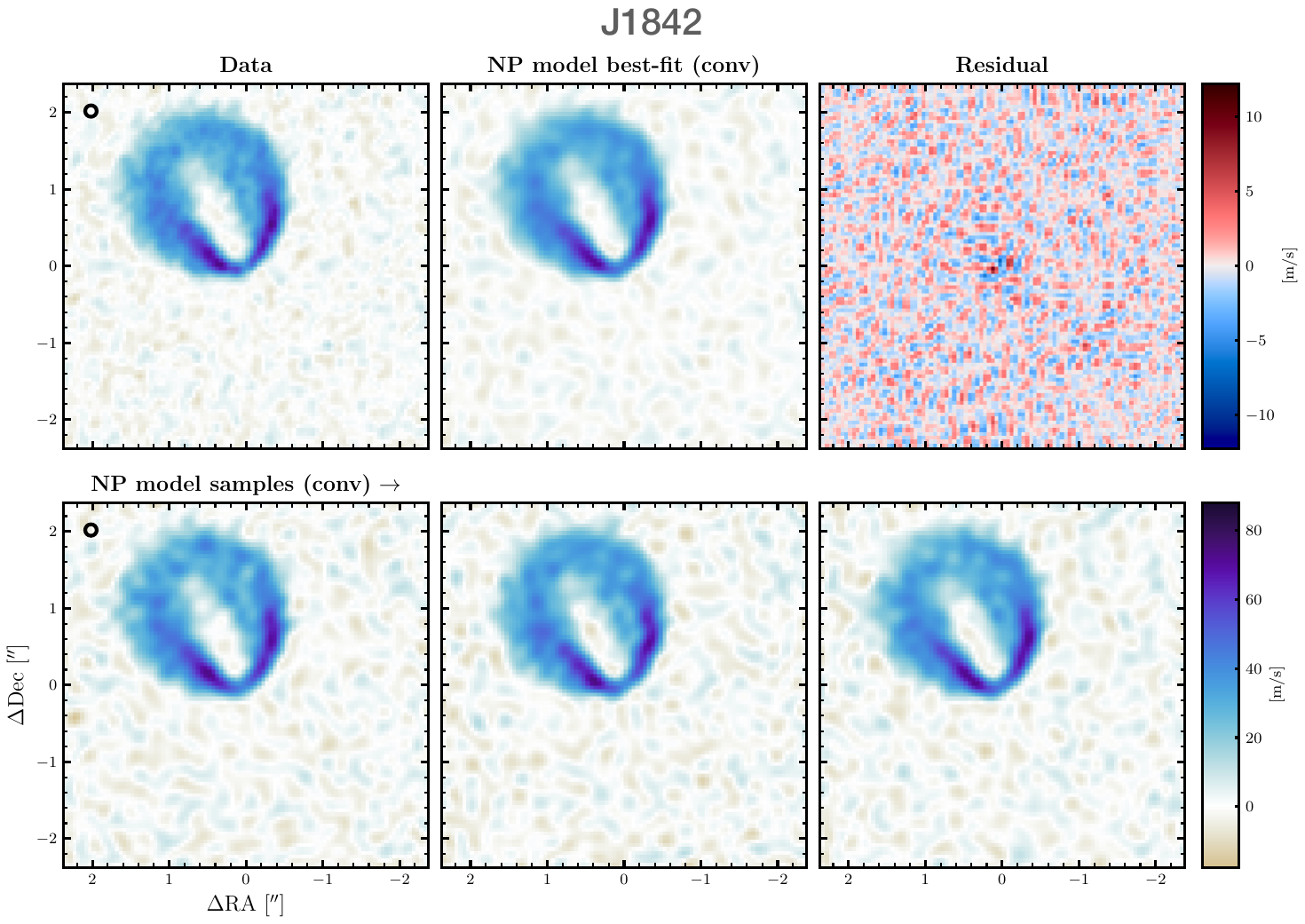}
    \caption{Single channel for the best-fitting non-parametric (NP) model and posterior samples for J1842. Top-left shows the data and beam. Top-middle shows the best-fitting model convolved with the beam. Top-right shows the residual between the data and best fit. Bottom row shows 3 posterior model channel samples convolved with the beam.}
    \label{fig:j1842}
\end{figure*}

In Figure~\ref{fig:j1842} we show the best-fitting non-parametric model to J1842 for one channel compared to the data in the top row.
The residuals are mostly noise.
The bottom row shows 3 copies of the same channel from different sample cubes drawn from the model posterior and convolved with the beam, to demonstrate the typical difference between samples.
The large and bright structures are constant across the samples, while the smaller and dimmer features show more variation.
30 such samples were used with \textsc{Discminer} for the combined approach outlined in Section~\ref{sec:methods_discminer}.

\begin{figure*}
    \centering
    \includegraphics[width=0.92\textwidth]{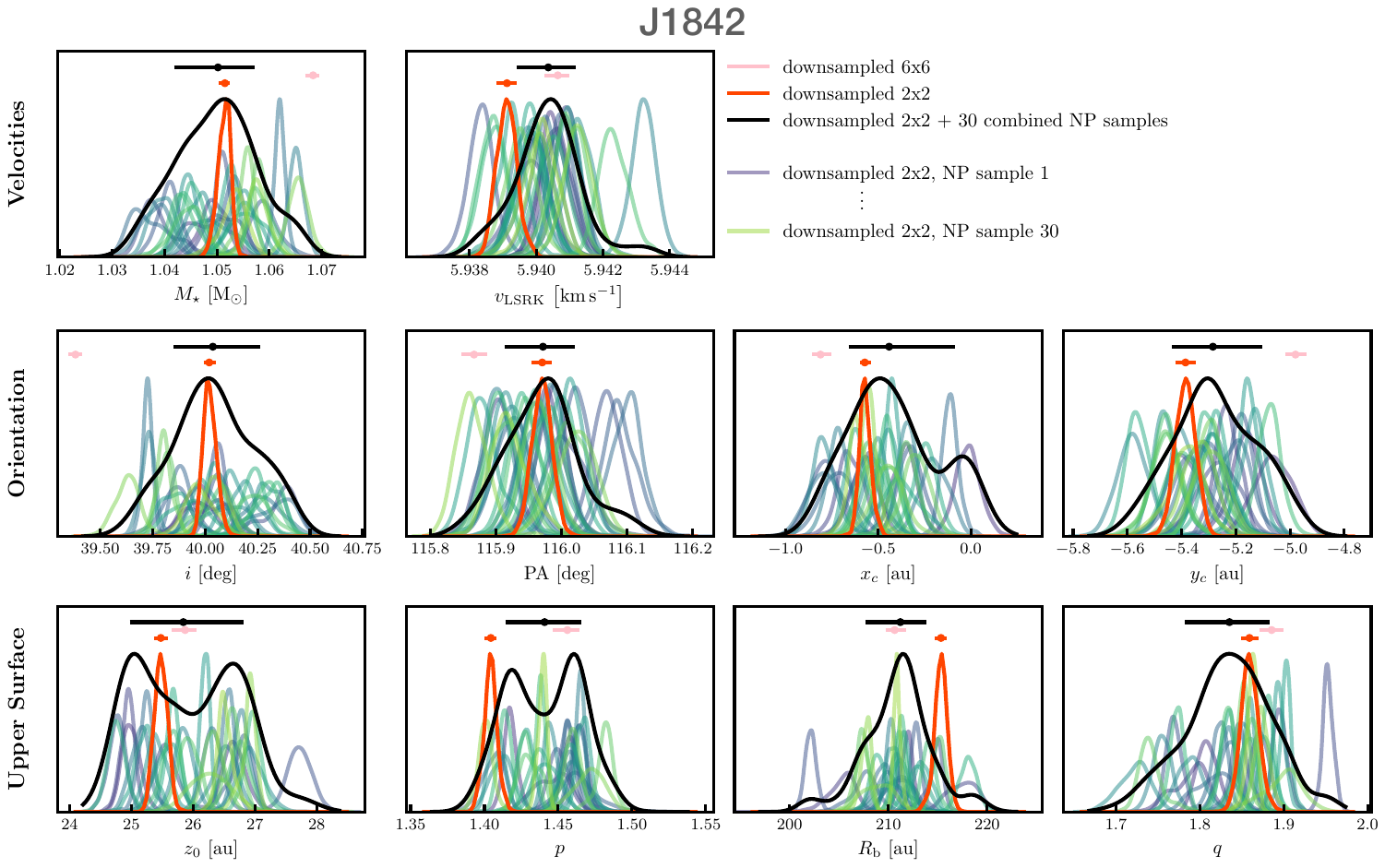}
    \caption{Posterior distributions for a subset of the J1842 \textsc{Discminer} model. Posteriors are plotted in each panel, with the corresponding median and $1\sigma$ quantiles indicated by the point and lines above the distributions. Fits from individual cube samples are shown in the blue-green colour map, and the combined distribution is shown in black. The red shows the posterior from fitting the data directly. The pink is taken from the production exoALMA model. The y-scale of the red and black curves have been scaled for easier comparison. The \textsc{Discminer} model parameters shown are stellar mass $M_\star$, systemic velocity $v_{\rm LSRK}$, inclination $i$, position angle ${\rm PA}$, model origin $(x_c, y_c)$, reference height $z_0$, height power-law index $p$, taper radius $R_{\rm b}$ and taper index $q$ \citep[see][]{Izquierdo_exoALMA}.}
    \label{fig:posteriors}
\end{figure*}

Figure ~\ref{fig:posteriors} shows the posterior distributions from the \textsc{Discminer} results.
We compare the distributions from running \textsc{Discminer} on the data directly (red) with the those from running on individual cube samples (blue-green colour map) as well as the combined distribution from all cube samples (black).
As outlined in Section~\ref{sec:methods_discminer}, this combined approach allows us to approximate the results that \textsc{Discminer} would find, were it to account for the spatially correlated noise.

We find that for all the parameters, the uncertainties are increased by approximately an order of magnitude.
The exception is the systemic velocity $v_{\rm LSRK}$, where the increase is only a factor of a few.

The posterior medians are not different between the data and combined approach to a statistically significant level, indicating that the non-parametric model is not causing a bias.
A few values are slightly offset, namely $v_{\rm LSRK}$, and the upper surface parameters $p$ and $R_{\rm b}$.
The differences in median surface parameters could be related to the fact that the parameters cause mostly spatially small changes to the shape of the channels, and so are potentially affected by accounting for the correlated noise.

Figure~\ref{fig:posteriors} also shows the posterior median and 1$\sigma$ quantiles for the production \textsc{Discminer} model, fit to the data with additional downsampling.
These estimates are mostly compatible with the others, except for $M_\star$ and $i$, which are offset to a significant level even accounting for the increased uncertainty from the correlated noise.
This could be caused by the extra downsampling inducing spurious smoothing, biasing the inclination low and the mass high to adjust.
While this bias is only on the order of one percent, it is statistically significant and is important to consider if the uncertainties from the analysis are to be used.

Since the combined posterior distributions shown are only from 30 cube samples, their overall shape is likely not accurately reconstructed, and so their non-Gaussian appearance in Figure~\ref{fig:posteriors} should not be taken too seriously, including the apparent bi-modality in $z_0$ and $p$.
However, we verified that the estimates of the medians and quantiles become stable at around 10--20 cube samples.

\section{Discussion} \label{sec:discussion}

We have demonstrated that our method for generating probabilistic moment maps is capable of debiasing line-of-sight velocity measurements associated with partially resolved substructures.
The correction is important for any studies looking to measure planet masses \citep{bollati2021}, disk masses \citep{veronesi2021,lodato2023,martire2024,speedie2024,Longarini_exoALMA} or vertical shear instability strength \citep{barrazaalfaro2021,Barraza_exoALMA} directly from the strength of velocity perturbations, as the bias would result in the underestimation of those quantities.
For example, measurements of disk mass using the so-called gravitational instability ``wiggle'' \citep{longarini2021,terry2022}, rely on comparing the deflection of the isovelocity curve near systemic to an analytic model.
As we have shown, these wiggles are systematically underestimated by conventional methods, but (at least partially) corrected for by our approach.
Analyses within exoALMA that rely on measuring velocity perturbations, for example the measurement of rotation curve deviations \citep{Stadler_exoALMA}, may be similarly affected, although a in-depth study is beyond the scope of this paper.
Our method is less useful in the limits where the substructures are either not resolved or well resolved.
As seen in \citet{Pinte_exoALMA}, the sample contains many partially resolved substructures.
The model is also likely to perform significantly worse for low signal-to-noise observations, although we did not test this.

The debiasing effect may also be important for measurements that rely on intensity, such as temperature and pressure, both of which exhibit substructures in the exoALMA sample \citep{Galloway_exoALMA,Stadler_exoALMA}.
Measures of turbulence may also be affected by the reduced line widths recovered with our method \citep{teague2016}, although forward models that fit to visibilities are not subject to bias from beam smearing \citep{flaherty2015,flaherty2020,Hardiman_exoALMA}.
The additional contrast we found in the line widths along the substructures may be useful for searches of embedded planets through line broadening \citep{Izquierdo_exoALMA}.

Additionally, our method provides uncertainties that take into account the spatially correlated nature of the noise.
Since many samples of each map are created, it is straightforward to propagate this uncertainty downstream into data products.
This could be used to quantify the correlation in adjacent data points along extracted rotation curves, as existing approaches are not able to take this into account \citep{Longarini_exoALMA}.

As alluded to in the Section~\ref{sec:overview}, the method is applicable to any measurement derived from the data.
For example, while we assumed Gaussian line profiles for the creation of moment maps, there is nothing about the method that requires this choice.
Recent improvements that more realistically characterise the line profile using higher order functions or multiple components \citep{teague2018a,Izquierdo_exoALMA} could be used in place of the Gaussians assumed here.
The only stipulation is that it must be computationally feasible to calculate the data product over many sample cubes.

Moreover, the non-parametric model itself makes few assumptions about the form of the data.
While we used the fiducial exoALMA $^{12}$CO images here, in principle the data may have any channel spacing or beam size and shape.
At full spectral resolution adjacent channels will be correlated \citep{loomis2018}, which we do not account for.
More broadly, the model may be applied to any image subject to a point spread function (PSF) and noise, provided that the PSF is known.

By combining our approach with \textsc{Discminer}, we estimated how the accuracy of the inferred model is affected when accounting for spatially correlated noise.
For J1842 the uncertainties are increased by an order of magnitude.
Despite this, $M_\star$ and $i$ are still constrained to within approximately 1\%.
Additionally, we found that it only takes around 10--20 cube samples for the uncertainty estimates to stabilise.
The computational cost for this calculation is greater than that involved in fitting \textsc{Discminer} to the data, but is still achievable in a few days on a 64 core node.
The cost is eased by the fact that all runs after the first can be initialised near the typical set, as the difference between the cube samples is small.
For studies concerned with reliable uncertainties on the outputs of \textsc{Discminer}, this approach is viable.
This uncertainty is purely statistical however, and if the data are more complicated than assumed in the model, there will be systematic uncertainty that is unaccounted for.
This is the case in exoALMA, as the \textsc{Discminer} model assumes Keplerian rotation when in reality there are deviations due to self-gravity, pressure support, and substructures \citep{Longarini_exoALMA,Stadler_exoALMA,Izquierdo_exoALMA}.
The combined approach could also be extended to calculate uncertainties on data products derived using \textsc{Discminer}, as each cube sample has its own associated inferred disk model.
This would allow for probalistic 2D residuals, which would aid in probing the robustness of features like doppler-flips \citep{perez2018a,perez2020,pinte2023b}, spirals \citep{teague2021,calcino2022,teague2022,izquierdo2023,Izquierdo_exoALMA} and vortices \citep{huang2018a,boehler2021,Wolfer_exoALMA} while accounting for both the correlated noise in the data as well as the uncertainty in the subtracted disk model.

The model presented in this paper has its own unique advantages, but it is similar to existing methods in the literature.
The model itself is of the true intensity distribution in each channel and so it is similar to CLEAN \citep{hogbom1974} or regularised maximal likelihood (RML) imaging \citep{akiyama2017,eht2019, zawadzki2023, Zawadzki_exoALMA}.
Unlike those methods, our model fits to the already imaged data and not the raw visibilities.
We argue that our method therefore constitutes image \emph{analysis} and not image \emph{creation} like in CLEAN or RML, although we admit this distinction is not sharply defined.
RML is likely to provide higher accuracy, as it utilises more information.
The benefit provided by our model is that it is linear and so the posterior distribution has an analytic form that is cheap to sample from.
While in principle it may be possible to sample from the RML posterior, it may prove intractable, even with samplers like Hamiltonian Monte Carlo that scale well to high dimensions \citep{duane1987, neal2011}.

\textsc{Frank} \citep{jennings2020} fits an axisymmetric model to visibilities directly.
Our methods share similarities in that both construct a model from a Fourier series (\textsc{Frank} uses a generalised Fourier series) and regularisation with a GP prior.
The most practical difference between our approaches is that our model is of the 2D intensity distribution while \textsc{Frank} fits for the radial intensity profile.
\textsc{Frank} also operates under the empircal Bayes paradigm where the covariance structure of the GP is non-parametric, whereas our regularisation is equivalent to an exponential covariance kernel.

\citet{dia2023} presented a method to perform image reconstruction in the framework of Bayesian inference, thus providing a posterior distribution over images.
Their method uses score-based priors calculated from a neural network trained on many images of galaxies.
This approach is promising but hinges on the choice of training set for the prior.
For their test case, continuum images of HD~143006 and AS~209 from the DSHARP program \citep{andrews2018,huang2018b}, the model produced biased fits due to the mismatch in prior.
Variational Bayesian methods such as flows using deep learning have also been shown to be capable of providing tractable sampling from the image posterior \citep{sun2021}, although these methods are formally approximate.

\subsection{Caveats}

We assumed that the beam is Gaussian.
This is true for the CLEAN beam, which is obtained by matching a Gaussian to the synthesised beam's central component.
The synthesised beam contains sidelobes, meaning that in the final image, the noise is not correlated exactly according to the CLEAN beam \citep{tsukui2022}.
We suspect that for the exoALMA data the impact on the analysis would be minor, due to the large number of baselines that ALMA has.
Parametric analyses on ALMA line emission data have found that switching from fitting in the image plane to complex visibilities results in negligible difference \citep{flaherty2018}.
It is unclear whether this generalises to our model however, due to the increased flexibility as compared to parametric approaches.

The impact of this on the model could be tested by processing the benchmark in a more realistic way with finite sampling of the uv-plane, but we leave this for future work.

The unconvolved model channels show boundary artefacts in the outermost couple of pixels.
This is caused by representing the beam convolution with a matrix multiplication, as this implicitly assumes the pixel values are zero outside the image.
This problem is minimised by ensuring that the images are cropped such that a few signal-free pixels are included along the edges.

With the implementation described here, the complexity of the model scales $\mathcal{O} (n^2 p + p^3)$, since we fit with numerical least squares (see Appendix~\ref{sec:numerical}).
This makes it prohibitive to use more pixels than around $100 \times 100$.
The memory required to store ${\bf A}$ also approaches a terabyte at around $125 \times 125$ pixels.
Conjugate gradient descent \citep[CGS;][]{hestenes1952} would provide a dramatic speed-up in solve time \citep[e.g.][]{shewchuk}, but does not solve the memory issue.
Methods that avoid the construction of ${\bf A}$, instead representing the system using linear operators \citep[][]{fong2011,pylops}, are a potential avenue for massive speed increases and memory requirement decreases.

As already touched on, our model fits to the CLEAN image and not the raw visibilities.
This is a drawback, as information is being discarded.
The model is also subject to biases or correlations induced through the CLEANing process.
Our approach hinges on the GP prior for the image itself, where the Fourier series is a computational trick more than a model of visibility space.
It is not immediately clear how this approach could be modified to fit visibilities instead.
A GP prior could be used for the visibilities, although a kernel that encourages smoothness would not be appropriate.
This kind of regularisation would be weaker than that used in successful approaches like RML, and so it is unclear whether or not it would be viable.

\section{Conclusions} \label{sec:conclusion}

We present a new method for generating probabilistic data products from spectral line data of protoplanetary disks, and apply it to a subset of the exoALMA sample.
Our main findings are summarised as follows:
\begin{enumerate}
    \item We model the data with a flexible, linear model of the image intensities prior to beam convolution, assuming spatially correlated noise.
    The model is composed of a Fourier basis, where modes with higher frequencies are increasingly suppressed.
    \item Model linearity results in an analytic posterior.
    This allows for the fast generation of sample cubes, which can be used to estimate the posterior distribution for any data product, taking into account the correlated noise in the data.
    \item As an example, we presented probabilistic  peak velocity, peak intensity, and line width maps of HD~135344B, CQ~Tau and MWC~758.
    We show that our method more accurately reconstructs line-of-sight velocities associated with substructures than with conventional methods, which underestimate localised perturbations due to beam smearing.
    Our method recovers perturbation amplitudes $50$--$150 \, {\rm ms^{-1}}$ larger than conventional methods.
    This debiasing allows more accurate estimates of planet masses, disk masses, vertical shear instability strength, and other analyses that rely on measuring deviations from Keplerian rotation.
    \item We also combined our approach with \textsc{Discminer}, using J1842 as a case study.
    We found that when taking into account the spatially correlated noise in the data, the uncertainties on the \textsc{Discminer} model parameters increase by an order of magnitude.
\end{enumerate}
The following data products will be publicly available
\begin{itemize}
    \item Cube samples for our fits to HD~135344B, CQ~Tau, MWC~758 and J1842.
    \item Probabilistic peak velocity, peak intensity, and line width maps for HD~135344B, CQ~Tau and MWC~758, including the best-fit, uncertainties, and individual samples.
\end{itemize}



\appendix

\section{Impact of hyperparameter selection} \label{sec:hyperparam}

As the hyperparameters $\lambda$ and $s$ appear directly in the expression for the model posterior mean (Eq.~\eqref{eq:Xhat}) and covariances (Eq.~\eqref{eq:Sigma}), the choice of their values will impact both the predictions and uncertainties in our results.
Increasing either $\lambda$ or $s$ in general leads to spatially smoother predictions and decreased variance in the model.
Setting $\lambda$ and $s$ to be too large will result in models that smear out the substructures we are interested in, while setting them too small will result in model images with too much high frequency noise.

Ideally, one would set the values of $\lambda$ and $s$ using the data themselves.
This can be achieved with cross-validation (CV), where the hyperparameters are tuned iteratively by partioning the data into testing and training sets and assessing the predictive performance on the held-out data \citep[e.g.][]{browne2000}.
For our method varying the hyperparameters does affect the model predictions but the convolved model predictions are altered only negligibly since the regularisation mostly impacts modes of higher frequencies.
Since the CV error metric would be between the data and the convolved model, this avenue does not provide a reliable way to ensure that the model prior to convolution is performant.

Alternatively, we could forward model a set of realistic, known images to create synthetic data which could then be used as a means of calibrating the prior and selecting hyperparameter values.
While an analysis of this type would strengthen confidence in the hyperparameter choice, a full implementation is out of scope for the immediate focus of this work, in which our priority is to introduce the core methodology.
Nonetheless, as demonstrated in the benchmark (Section~\ref{sec:benchmark}), our chosen hyperparameters yield plausible reconstructions of the intensity gradients around substructures leading to more accurate inferences when compared with the ground truth.

\begin{figure*}
    \centering
    \includegraphics[width=0.9\textwidth]{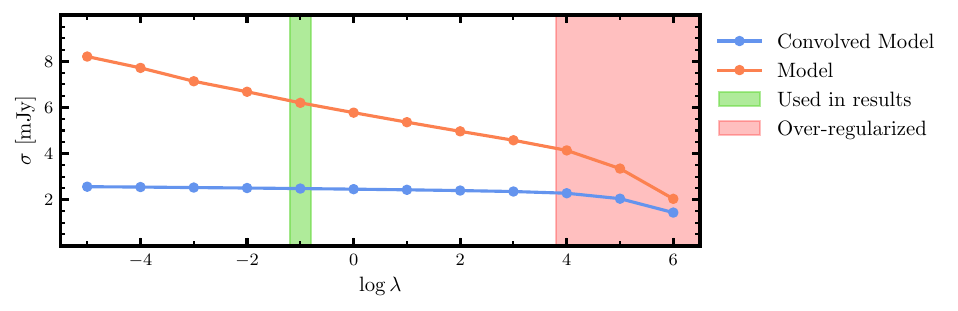}
    \caption{Uncertainty in model prediction as a function of regularisation strength $\lambda$ for the model before (orange) and after (blue) convolution with the beam. The green region indicates the value of $\lambda$ used in our results. The red region indicates the region where the model is clearly over-regularised.}
    \label{fig:sigma_lambda}
\end{figure*}

We should also be concerned with the impact of the hyperparameter choices on the uncertainties in our results.
To assess this, we calculated $\Sigma$ for a large range of possible $\lambda$ values for the fit to the HD~135344B data.
We varied only $\lambda$ and kept $s=0.1$ fixed because it has been found that for this kernel only the product $\lambda^2 \sqrt{s}$ is actually identifiable from data in general \citep{zhang2004,gelman2017}.
We then found the marginal standard deviation on the model predictions for each pixel before and after convolution as

\begin{align}
    \sigma \left( {\bf I} \right) &= \sqrt{{\rm diag} \left({\bf A}_{xy} \boldsymbol{\Sigma} {\bf A}_{xy}^{\top} \right)}, \\
    \sigma \left( {\bf I}_{\rm conv} \right) &= \sqrt{{\rm diag} \left( {\bf A} \boldsymbol{\Sigma} {\bf A}^{\top} \right)},
\end{align}

where ${\rm diag}$ extracts the diagonal of a matrix as a vector.
We note that the calculated variance is almost identical in each pixel for a given $\lambda$.
The resultant values are shown in Figure~\ref{fig:sigma_lambda}.
We see the model prediction uncertainty decreases approximately linearly with $\log \lambda$, while the convolved model uncertainty is essentially flat.
At $\lambda \approx 10^4$ there is a turn-off from this behaviour where the model uncertainty decreases more rapidly and the convolved model uncertainty begins to decrease as well.
This difference is because the regularisation acts in the model to suppress high frequency modes: as the prior becomes weaker the model can contain more high frequency noise, but the convolved model is basically unchanged as the high frequency information doesn't survive the convolution.
At very large $\lambda$, shaded red in the Figure, the regularisation is strong enough to suppress modes of lower frequencies such that the variance in the convolved model predictions is affected.
This corresponds to a prior so strong that the convolved model predictions will be smoother than the beam and so we label this as over-regularised.

The green shaded area in Figure~\ref{fig:sigma_lambda} corresponds to the level of regularisation we used in our results, $\lambda = 10^{-1}$.
The downstream data products presented in Section~\ref{sec:results_moments} depend on per-pixel quantities, and so their uncertainty will also be underestimated if our regularisation is too strong.
This will likely be a minor effect, as a decrease in regularisation of 4 orders of magnitude only increases the uncertainty by a factor of around 30\%.
For the analysis using \textsc{discminer} presented in Section~\ref{sec:results_discminer}, we used samples from the convolved model, and so the results and uncertainties should be unaffected by a different choice of hyperparameters unless over-regularised.

\section{Numerical considerations} \label{sec:numerical}

In Section~\ref{sec:linear_algebra} when outlining our model we gave the solution for ${\bf \hat{X}}$ as Eq.~\eqref{eq:Xhat}.
In practice we instead solve for the equivalent expression \citep{henderson1981} when $\boldsymbol{\mu} = {\bf 0}$
\begin{align}
    {\bf \hat{X}} &= \boldsymbol{\Lambda} {\bf A}^\top \left( {\bf A} \boldsymbol{\Lambda} {\bf A}^\top + {\bf C} \right)^{-1} {\bf I}_{\rm data}, \label{eq:Xhat_calc_overparam}
\end{align}
since it is faster and more numerically stable for the case where $p>n$ \citep[e.g.][]{hoggvillar} which we used in this work.
It also avoids inverting ${\bf C}$.
The solution in Eq.~\eqref{eq:Xhat} is faster and more numerically stable if $n<p$, and we left it in the main text since its connection to the regular GLS solution in Eq.~\eqref{eq:GLS} is more obvious.

Additionally, the condition number of ${\bf H}$, and so also ${\bf A}$ and ${\bf C}$, can be very large due to the small tails of the Gaussian distribution used for the beam kernel $h$ in Eq.~\eqref{eq:conv_matrix}.
This problem worsens when the beam is more resolved, i.e. when the number of pixels per beam is increased.
We deal with this issue in two ways.
First, we downsample the cubes spatially by a factor of 2 which reduces the number of pixels per beam, but not so much so that the correlations between nearby pixels is no longer resolved.
This also helps reduce computation cost since it reduces $n$ by a factor of 4.
Second, following the advice of \citet{hoggvillar}, we avoid directly calculating the inverse of matrices (with the exception of $\boldsymbol{\Lambda}^{-1}$ since it is diagonal). 
We instead solve for products of the inverse using numerical least squares with \textsc{NumPy}'s \codesnip{linalg.lstsq} function with \textsc{rcond} set to $10^{-16}$ to zero out small eigenvalues near machine precision.
That is, for some system ${\bf A} {\bf X} = {\bf B}$ we use \codesnip{linalg.lstsq(A, B)} to find ${\bf X}$ instead of ever calculating ${\bf A}^{-1}$.
This approach is especially important for calculating the uncertainties on ${\bf \hat{X}}$ since it \emph{requires} the inverse of ${\bf C}$.
We break the solution in Eq.~\eqref{eq:Sigma} into the following steps.
We set $\boldsymbol{\Gamma} = {\bf A}^\top {\bf C}^{-1}$ giving $\boldsymbol{\Gamma} {\bf C} = {\bf A}^\top$ where we can then solve for $\boldsymbol{\Gamma}$ using the above approach.
We then calculate the expression inside the brackets with $\boldsymbol{\Gamma} {\bf A} + \boldsymbol{\Lambda}^{-1}$ = ${\bf A}^\top {\bf C}^{-1} {\bf A} + \boldsymbol{\Lambda}^{-1}$.
Finally, we use a LU-decomposition for the final unavoidable direct inversion, yielding $\boldsymbol{\Sigma}$.

After calculating ${\bf \hat{X}}$ and $\boldsymbol{\Sigma}$ we need to draw samples from the posterior distribution given by Eq.~\eqref{eq:posterior} which is an $p$-dimensional normal distribution.
Let ${\bf Z}$ be a vector containing $p$ independent draws from the standard normal distribution (zero mean and unit variance), and let ${\bf B}$ be any real matrix such that $\boldsymbol{\Sigma} = {\bf B} {\bf B}^\top$.
We can then generate samples ${\bf X}$ from the desired distribution using
\begin{align}
    {\bf X} = {\bf \hat{X}} + {\bf B} {\bf Z}.
\end{align}
In theory $\boldsymbol{\Sigma}$ is guaranteed to be symmetric, and it may be positive definite or positive semi-definite.
In practice $\boldsymbol{\Sigma}$ is often not quite symmetric due to numerical precision and so we calculated a symmetric version $\boldsymbol{\Sigma}' = 0.5 \left( \boldsymbol{\Sigma} + \boldsymbol{\Sigma}^\top \right)$ which we use to find ${\bf B}$.
If $\boldsymbol{\Sigma}$ is positive definite then we use a Cholesky decomposition
\begin{align}
    \boldsymbol{\Sigma} = {\bf L} {\bf L}^\top
\end{align}
where ${\bf L}$ is a real, lower-triangular matrix.
If $\boldsymbol{\Sigma}$ is only positive semi-definite we instead use a spectral decomposition
\begin{align}
    \boldsymbol{\Sigma} = {\bf Q} {\bf D} {\bf Q}^\top = ({\bf Q} {\bf D}^{1/2}) ({\bf Q} {\bf D}^{1/2})^\top
\end{align}
where ${\bf Q}$ is an orthogonal matrix containing the eigenvectors of $\boldsymbol{\Sigma}$ as columns and ${\bf D}$ is a diagonal matrix whose diagonal elements are the corresponding eigenvalues, which are all nonnegative.

\section{Benchmark model} \label{sec:bench_model}

The radial intensity profile we use is
\begin{align}
    I(r) =
    \begin{cases} 
        \left[0.35 \, \exp{\left( - \left[\frac{r}{40 \, {\rm au}} \right]^{0.45} \right)} + 0.5 \, \exp{\left( - \frac{1}{2} \left[\frac{r - 7 \, {\rm au}}{15 \, {\rm au}} \right]^{2} \right)} \right] \, {\rm Jy/beam} & r \leq 220 \, {\rm au}, \\
        0 \, {\rm Jy/beam} & {\rm elsewhere}, \\
     \end{cases}
\end{align}
and the line width profile is
\begin{align}
    w(r) = 0.65 \exp{\left( - \left[\frac{r}{500 \, {\rm au}} \right]^{0.2} \right)} \, {\rm km/s}.
\end{align}
We assume azimuthal, Keplerian motions $v(r) = v_{\phi}(r)$, with 3 Gaussian ring perturbations placed at different radii
\begin{align}
    v_{\phi} (r) &= \left[1 - 0.25 \, G_{70}(r) + 0.35 \, G_{90}(r) + 0.25 \, G_{140}(r) \right] v_{\rm K} (r), \\
    v_{\rm K} (r) &= \frac{G M_{\star}}{r}, \\
    G_{R}(r) &= \exp{\left(- \frac{1}{2} \left[ \frac{r - R \, {\rm au}}{10 \, {\rm au}} \right]^2 \right)},
\end{align}
such that the perturbations have amplitudes of $-0.25$, $0.35$ and $0.25$ of Keplerian at radii $70$, $90$ and $140 \, {\rm au}$ respectively.
The intensity and line width radial profiles were chosen by hand to produce roughly similar channels to HD~135344B and their exact form should not be treated as a careful fit to the data.

\vspace{5mm}
\software{Astropy \citep{astropycollaboration2022a}, NumPy \citep{harris2020}, SciPy \citep{virtanen2020}, matplotlib \citep{hunter2007}, cmasher \citep{vandervelden2020}}.
\bibliography{refs.bib}{}
\bibliographystyle{aasjournal}

\section*{Acknowledgments}

This paper makes use of the following ALMA data: ADS/JAO.ALMA\#2021.1.01123.L. ALMA is a partnership of ESO (representing its member states), NSF (USA) and NINS (Japan), together with NRC (Canada), MOST and ASIAA (Taiwan), and KASI (Republic of Korea), in cooperation with the Republic of Chile. The Joint ALMA Observatory is operated by ESO, AUI/NRAO and NAOJ. The National Radio Astronomy Observatory is a facility of the National Science Foundation operated under cooperative agreement by Associated Universities, Inc. We thank the North American ALMA Science Center (NAASC) for their generous support including providing computing facilities and financial support for student attendance at workshops and publications.
This work was performed in part on the OzSTAR national facility at Swinburne University of Technology. The OzSTAR program receives funding in part from the Astronomy National Collaborative Research Infrastructure Strategy (NCRIS) allocation provided by the Australian Government, and from the Victorian Higher Education State Investment Fund (VHESIF) provided by the Victorian Government. 
TH, CH and IH are supporteded by Australian Government Research Training Program (RTP) Scholarships.
JB acknowledges support from NASA XRP grant No. 80NSSC23K1312. MB, DF, JS have received funding from the European Research Council (ERC) under the European Union's Horizon 2020 research and innovation programme (PROTOPLANETS, grant agreement No. 101002188). Computations by JS have been performed on the `Mesocentre SIGAMM' machine, hosted by Observatoire de la Cote d'Azur. PC acknowledges support by the Italian Ministero dell'Istruzione, Universit\`a e Ricerca through the grant Progetti Premiali 2012 – iALMA (CUP C52I13000140001) and by the ANID BASAL project FB210003. SF is funded by the European Union (ERC, UNVEIL, 101076613), and acknowledges financial contribution from PRIN-MUR 2022YP5ACE. MF is supported by a Grant-in-Aid from the Japan Society for the Promotion of Science (KAKENHI: No. JP22H01274). JDI acknowledges support from an STFC Ernest Rutherford Fellowship (ST/W004119/1) and a University Academic Fellowship from the University of Leeds. Support for AFI was provided by NASA through the NASA Hubble Fellowship grant No. HST-HF2-51532.001-A awarded by the Space Telescope Science Institute, which is operated by the Association of Universities for Research in Astronomy, Inc., for NASA, under contract NAS5-26555. CL has received funding from the European Union's Horizon 2020 research and innovation program under the Marie Sklodowska-Curie grant agreement No. 823823 (DUSTBUSTERS) and by the UK Science and Technology research Council (STFC) via the consolidated grant ST/W000997/1. CP and DP acknowledge Australian Research Council funding via FT170100040, DP18010423, DP220103767, and DP240103290. GR acknowledges funding from the Fondazione Cariplo, grant no. 2022-1217, and the European Research Council (ERC) under the European Union's Horizon Europe Research \& Innovation Programme under grant agreement no. 101039651 (DiscEvol). H-WY acknowledges support from National Science and Technology Council (NSTC) in Taiwan through grant NSTC 113-2112-M-001-035- and from the Academia Sinica Career Development Award (AS-CDA-111-M03). GWF acknowledges support from the European Research Council (ERC) under the European Union Horizon 2020 research and innovation program (Grant agreement no. 815559 (MHDiscs)). GWF was granted access to the HPC resources of IDRIS under the allocation A0120402231 made by GENCI. Support for BZ was provided by The Brinson Foundation. Views and opinions expressed by ERC-funded scientists are however those of the author(s) only and do not necessarily reflect those of the European Union or the European Research Council. Neither the European Union nor the granting authority can be held responsible for them.

\end{document}